\documentclass[twocolumn,pra,preprintnumbers,superscriptaddress]{revtex4-2}

\usepackage{dsfont, xcolor, comment}
\usepackage{graphicx}% Include figure files
\usepackage{dcolumn}% Align table columns on decimal point
\usepackage{bm}% bold math
\usepackage{amsmath}
\usepackage{svg}
\usepackage{import}
\usepackage{stmaryrd}
\usepackage{pgf}
\usepackage{braket}
\usepackage{amsfonts}
\usepackage{titletoc}

\makeatletter
\def\maketitle{
\@author@finish
\title@column\titleblock@produce
\suppressfloats[t]}
\makeatother

\begin{document}

\title{Quantum manifestations of converse symmetry breaking}
\author{Daniel P. Buser}
\author{Biswabibek Bandyopadhyay}
\affiliation{\footnotesize Department of Physics and Astronomy, Northwestern University, Evanston, IL 60208, USA}
\affiliation{\footnotesize  Center for Network Dynamics, Northwestern University, Evanston, IL 60208, USA}

\author{Adilson E. Motter}
\affiliation{\footnotesize Department of Physics and Astronomy, Northwestern University, Evanston, IL 60208, USA}
\affiliation{\footnotesize Center for Network Dynamics, Northwestern University, Evanston, IL 60208, USA}
\affiliation{\footnotesize Department of Engineering Sciences and Applied Mathematics, Northwestern University, Evanston, IL 60208, USA}
\affiliation{\footnotesize Northwestern Institute on Complex Systems, Northwestern University, Evanston, IL 60208, USA}
\affiliation{\footnotesize Institute for Quantum Information Research and Engineering, Northwestern University, Evanston, IL 60208, USA}
%
%\date{\today}
%
\begin{abstract}
Quantum synchronization holds promise in advancing quantum technologies such as sensing, communication, and computation. A key challenge in this rapidly developing field is to identify mechanisms that can promote the emergence of synchronous dynamics. Here, we show that quantum synchronization can be enhanced by introducing controlled heterogeneity in the form of coupling mismatches. 
Using spin van der Pol oscillators as a model system, we show that this mechanism constitutes a quantum analog of the phenomenon of converse symmetry breaking previously identified in classical oscillator networks, 
where a synchronous 
state is stabilized by explicit symmetry breaking. We quantify how quantum effects modify this behavior by tuning the degree of quantumness through variations of the spin quantum number. These findings suggest that quantum synchronization can be enhanced in experiments by harnessing rather than suppressing heterogeneity.
\end{abstract}

\maketitle

\section{Introduction}

Synchronization is a phenomenon traditionally studied in classical dynamics~\cite{sync}, ranging from the coordinated firing of neurons and frequency-synchronized operation of power grids to the rhythmic flashing of fireflies. 
More recently, this framework has been extended to open quantum systems~\cite{Bianconi_review}, including quantum entrainment~\cite{Lee_prl, Bruder_prl1, squeezing} and quantum mutual synchronization under various coupling schemes~\cite{Lee_prl, lee_pre, Bruder_annalender, dirk2017, quadratic_qsync}. 
A major focus of this emerging area is quantum limit-cycle systems, such as quantum van der Pol (vdP) oscillators~\cite{Lee_prl, Bruder_prl1},
recently implemented in trapped-ion experiments~\cite{qsync_expt1, qsync_expt2}.
Quantum synchronization associated with limit cycles has also been explored in spin-based systems, both theoretically~\cite{BruderSmall, spin1_entanglement, optimal_spin_qsync, macroscopic_qsync} and experimentally~\cite{spin1_expt, IBMQspin1}.

Previous research has shown that quantum synchronization can be influenced by various factors, including nonlinearity~\cite{relaxation_qsync, nonlinearity_qsync}, noise~\cite{noise_qsync_2016, noise_qsync_2022, noise_qsync_expt}, and measurement~\cite{measurement_Kato, measurement_kwek, measurement_bruder}.
Another recurrent factor is heterogeneity, which arises naturally in laboratory settings due to mismatches in oscillator and coupling parameters.
Although heterogeneity is generally viewed as detrimental,
research in classical systems has shown that
it can often be harnessed to enhance synchronization \cite{sagues2007spatiotemporal} in both driven~\cite{braiman1995taming,tessone2006diversity} and  self-sustained~\cite{montaseri2016diversity,punetha2022how} oscillators. 
Converse symmetry breaking (CSB) epitomizes this idea by leveraging intrinsic heterogeneity to stabilize synchronous states~\cite{AEMprl2016,csb_molnar}.
In classical systems, CSB occurs when an otherwise unstable synchronous state---which is invariant under node permutations and thus symmetric---is stabilized by explicitly breaking the corresponding permutation symmetry of the system.  
This can be achieved through heterogeneity in coupling  structure~\cite{rajroy_motter,zhang2022isochronous}, oscillator parameters~\cite{Zhang2021, csb_perego},  spatiotemporal noise~\cite{coherent_kiss_motter, enhance_sync_noise}, other forms of external driving~\cite{brandt2006synchronization}, or
combinations of these~\cite{nair2021using}, with  emerging applications across natural and engineered systems~\cite{yang2022emergent, maltsev2022functional, arthur2025optimal,dahmen2026how}.
The influence of heterogeneity on quantum synchronization has been examined in Refs.~\cite{Zambrini2013, qsync_blockade, asymm_induced_qsync} via frequency or modulation mismatches, and in Refs.~\cite{Zambrini_corr, Restrepo_dipoles} through the coupling.
Importantly,  Refs.~\cite{qsync_blockade, blockade_nigg} report the unique phenomenon of quantum synchronization blockade, 
in which the synchronization of anharmonic oscillators is inhibited when their natural frequencies are identical and facilitated when they are detuned, in contrast to the corresponding classical systems.
More recently, the concept of synchronization blockade has been extended to systems exhibiting destructive interference~\cite{interference_blockade} and nonreciprocal coupling~\cite{nonreciprocal_blockade}.
Despite these advances, leveraging heterogeneity to promote synchronization in the spirit of CSB has not yet been explored in quantum systems.

In this article, we address this gap by introducing controlled heterogeneity through coupling mismatches in a system of quantum spin vdP oscillators. 
Originally introduced in Ref.~\cite{kato2025}, these oscillators are ideal candidates 
for comparing
quantum 
and 
classical synchronization, 
since the transition between these regimes can be achieved by varying the spin quantum number, as we show here. 
We demonstrate that controlled coupling mismatches can enhance quantum synchronization in a network of spin vdP oscillators.
We further quantify how the dynamics evolve with the system’s quantumness by comparing the dynamics of the quantum system (low spin) with those in the classical limit (infinite spin) using the same synchronization measure. Our results link CSB of the classical system to the synchronization enhancement induced by
controlled heterogeneity in the deep-quantum regime, establishing a direct relationship between this classical synchronization mechanism and its quantum counterpart.

\section{Uncoupled spin \lowercase{vd}P oscillator}

Unlike standard quantum vdP oscillators~\cite{Lee_prl, Bruder_prl1}, spin vdP oscillators have an inherently finite-dimensional Hilbert space, allowing one to probe the full space for genuine quantum nonlinear effects where level quantization and quantum noise play dominant roles. Spin vdP oscillators can be experimentally realized in various established platforms, including trapped ions~\cite{trapped_ions}, nitrogen-vacancy centers~\cite{NV_centers}, and superconducting artificial atoms~\cite{superconducting_artificial_atoms}, where three-level systems have been demonstrated, making the spin vdP model directly relevant to current quantum technologies.

For a single spin vdP oscillator, we adopt the master equation introduced in Ref.~\cite{kato2025}
with the modification that the ladder operators $S^+$ and $S^-$ are swapped. This choice ensures that, in the classical limit, the model reduces to the Landau–Lifshitz equation for a spin precessing at a frequency $\omega$ in a magnetic field, subject to linear gain and nonlinear damping~\cite{damped_spin}. The master equation for a spin-$S$ system (with $\hbar=1$) then takes the form 
\begin{equation}\label{eq:single_qspin}
    \dot{\rho} = -{\rm i}\,[\omega S^z, \rho] + \gamma_\text{g} \mathcal{D}[S'^-]\rho + \frac{\gamma_\text{d}}{2S}\mathcal{D}[(S'^+)^2]\rho,
\end{equation}
where
$\rho$ is the density matrix, ${S'}^\pm = S^\pm/\sqrt{S}$
denote reduced ladder operators, and $S^z$ and $S^\pm$ obey the  spin operator algebra with commutators $[S^z,S^\pm]=\pm S^\pm$ and $[S^+, S^-] = 2S^z$. The superoperator $\mathcal{D}[K]\rho = K\rho K^\dagger - \frac{1}{2}\{K^\dagger K, \rho\}$ is the single-operator Lindblad dissipator~\cite{open_systems}.  The parameters $\gamma_\text{g}$ and $\gamma_\text{d}$ are the gain and damping rates, respectively. Note that $S=1$ is the smallest spin value for which the model is valid---and for which synchronization is possible \cite{BruderSmall}---since for spin-$\frac{1}{2}$ systems $(S^+)^2=0$.
The magnetic field is assumed to be in the $z$ direction, indicated by the unit vector $\hat{z}$.

\begin{figure}
    \centering
    \includegraphics[scale=0.18]{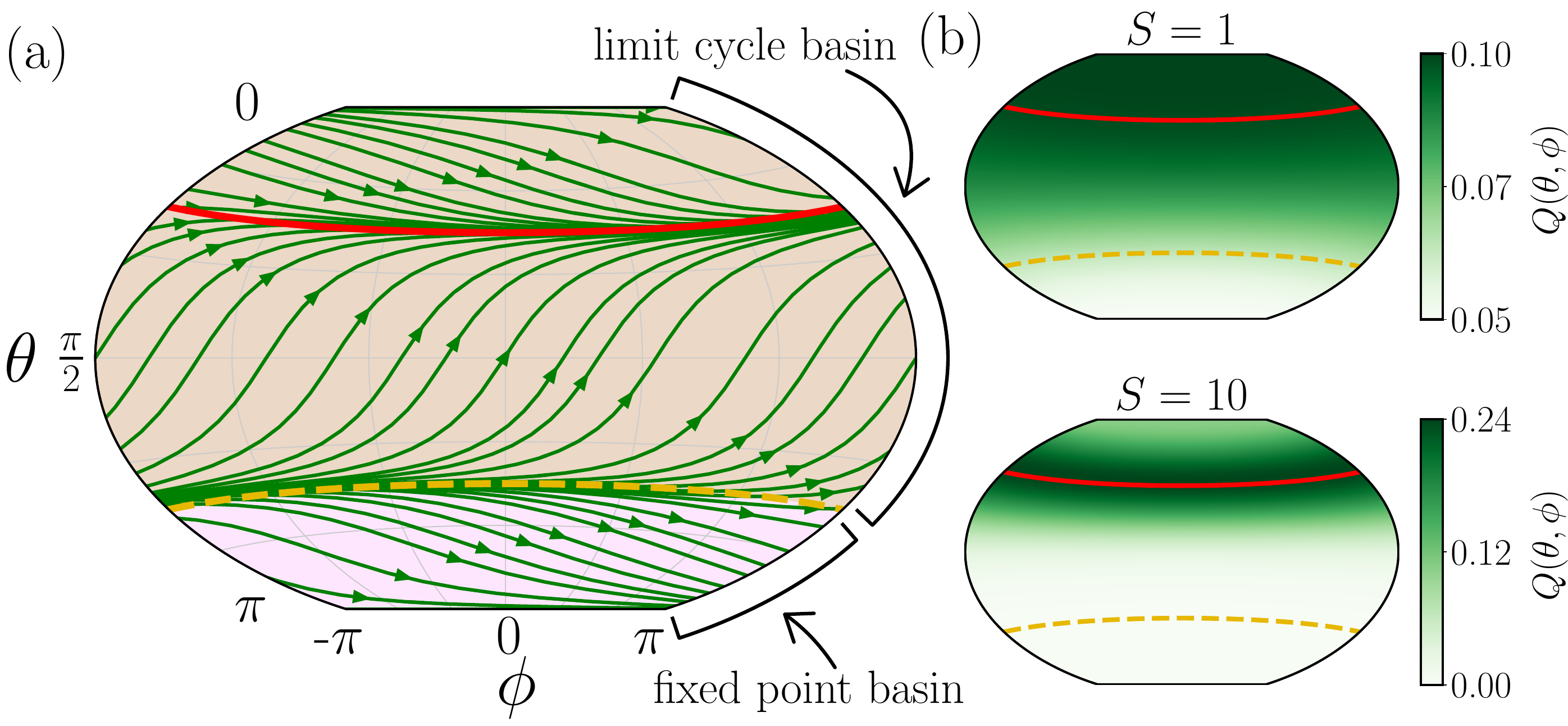}
    \vspace{-1mm}
    \caption{Dynamics of isolated classical and quantum spin vdP oscillators. 
    (a) Classical dynamics described by Eq.~\eqref{eq:classical_model}, exhibiting a stable cycle  at $\theta = \arcsin(\sqrt{\gamma_\text{g}/\gamma_\text{d}})$  (red solid line), an unstable cycle at $\pi - \arcsin(\sqrt{\gamma_\text{g}/\gamma_\text{d}})$ (yellow dashed line), and fixed points at  $\theta = 0$ (unstable)  and $\pi$ (stable).
    (b) Steady-state Husimi $Q$-function for the quantum model in Eq.~\eqref{eq:single_qspin} for $S=1$ (top) and $10$ (bottom). Both panels are for $\omega = 1$, $\gamma_\text{g} = 1$, and $\gamma_\text{d} = 2$ and show the spherical surface using the Winkel Tripel projection. 
    }
    \label{fig:limit_cycles}
\end{figure}

To investigate the behavior of the quantum model in the classical limit, we derive the classical equivalent of Eq.~\eqref{eq:single_qspin} by defining the classical angular momentum as $\vec{L} = \langle\vec{S}\rangle$, where $\vec{S} = (S^x, S^y, S^z)$ and $\langle\cdot\rangle = \text{Tr}(\rho\;\cdot)$ denotes the expectation value with respect to $\rho$. The dynamical equation for $\vec{L}$ can be found by using $\dot{\vec{L}}=\text{Tr}(\dot{\rho}\vec{S})$ while neglecting terms of order $O(\hbar)$ [one must first restore the factors of $\hbar$ in Eq.~\eqref{eq:single_qspin}] and approximating $\langle K_1K_2\rangle \approx \langle K_1 \rangle \langle K_2 \rangle$ for arbitrary operators $K_1$ and $K_2$. This leads to
\begin{align}
    \dot{\vec{L}} = \vec{L} \times \! \left[-\omega\hat{z} + \vec{L} \times \hat{z} \Big[ \frac{\gamma_\text{g}}{S} - \frac{\gamma_\text{d}}{S^3}\Big( S^2 - \big( \vec{L} \cdot \hat{z} \big)^2 \Big) \Big] \right],
    \label{eq:classical_model_vec}
\end{align}
which is the Landau-Lifshitz equation for damped gyromagnetic precession about the $z$ axis with additional nonlinear damping.
The spin quantum number $S$ corresponding to the norm of $\vec{L}$ is a conserved 
tunable quantity in the quantum model, which is a property consistently reflected in the classical model since $\dot{\vec{L}} \cdot \vec{L} = 0$.

Performing a change of variables to angular coordinates, the dynamical equations of the non-conserved quantities in Eq.~\eqref{eq:classical_model_vec}  become
\begin{subequations}\label{eq:classical_model}
    \begin{equation}\label{subeq:theta_eq}
        \dot{\theta} = \sin(\theta)\left[ \gamma_\text{g} - \gamma_\text{d} \sin^2(\theta) \right],
    \end{equation}
    \vspace{-0.7cm}
    \begin{equation}\label{subeq:phi_eq}
        \dot{\phi} = \omega,
    \end{equation}
\end{subequations}
where $0\leq\theta\leq\pi$ and $-\pi\leq\phi<\pi$. In the small-$\theta$ approximation, one has  $\sin(\theta)\approx\theta$ and the system can be interpreted as a variant of the Stuart-Landau oscillator model in polar coordinates with $\theta$ serving as the oscillation amplitude and $\phi$ as its phase~\cite{Lee_prl}.  Equation~\eqref{subeq:theta_eq} admits four steady-state solutions, of which two define limit cycles of constant $\theta$ on the sphere---one stable and one unstable.
The ratio $\gamma_\text{g}/\gamma_\text{d}$ controls the angle $\theta$ of the stable (unstable) limit cycle, with larger ratios corresponding to larger (smaller) angles at which the spin will precess. The remaining two solutions are at the poles of the sphere, corresponding to an unstable ($\theta = 0$) and a stable ($\theta = \pi$) fixed point. The two limit cycles merge at the equator when $\gamma_\text{g} = \gamma_\text{d}$, and the system becomes monostable at $\theta = \pi$ for $\gamma_\text{g} > \gamma_\text{d}$. Figure~\ref{fig:limit_cycles}(a) shows the vector field and limit cycles of the classical model in Eq.~\eqref{eq:classical_model}, while Fig.~\ref{fig:limit_cycles}(b) shows the Husimi $Q$-function of the steady states of the quantum model in Eq.~\eqref{eq:single_qspin}.
It follows that the regions of large Husimi $Q$-function weight tend to localize around the stable classical limit cycle with no phase preference, and that the degree of localization increases with  $S$.

\section{Coupled spin \lowercase{vd}P oscillators}

In our analysis of $N$ coupled spin vdP oscillators, we consider the model 
\begin{align}
    \nonumber\dot{\rho} &= \sum_{n=1}^N \Big(-{\rm i}\,[\omega S_n^z, \rho] + \gamma_\text{g} \mathcal{D}[{S'_n}^-]\rho + \frac{\gamma_\text{d}}{2S}\mathcal{D}[({S'_n}^+)^2]\rho\Big) \\ \nonumber
    &+ \sum_{n=1}^N\sum_{m=1}^N \bigg(\frac{1}{2}C_{nm}^a[{S'_n}^-{S'_m}^+, \rho] \\ 
    &+ C_{nm}^{h+}\mathcal{D}[{S'_n}^-, {S'_m}^+]\rho - C_{nm}^{h-^*}\mathcal{D}[{S'_n}^+, {S'_m}^-]\rho\bigg),
    \label{eq:master_equation}
\end{align} 
where $n$ indexes the individual spins, $\mathcal{D}[K_1, K_2]\rho = K_1\rho K_2 - 1/2\{K_2 K_1, \rho\}$ denotes the two-operator Lindblad dissipator, $\mathbf{C}^a$ represents the coherent contribution of the coupling,  $\mathbf{C}^{h\pm}$ represent the incoherent contribution, and an asterisk indicates the complex conjugate. The matrix $\mathbf{C}^a$ is anti-Hermitian and modifies the total Hamiltonian, while matrices $\mathbf{C}^{h+}$ and $\mathbf{C}^{h-}$ are Hermitian and are respectively positive semidefinite and negative semidefinite.  This coupling scheme is designed to yield a system that, in the classical regime, is capable of exhibiting CSB (as described below).

In model \eqref{eq:master_equation},
all  spins are identical in terms of their internal parameters $S$, $\omega$, $\gamma_\text{g}$, and $\gamma_\text{d}$. The symmetry of the system is therefore defined by the choice of couplings between the spins, which is determined by the matrices $\mathbf{C}^a$, $\mathbf{C}^{h+}$, and $\mathbf{C}^{h-}$.
These matrices can be summed into a single (complex) matrix $\mathbf{C}$, which is related to the adjacency matrices in the classical case. In our design of the network, we first construct $\mathbf{C}$, which in general can be any complex $N\times N$ matrix with arbitrary entries, and then decompose $\mathbf{C}$ into $\mathbf{C}^a$, $\mathbf{C}^{h+}$, and $\mathbf{C}^{h-}$,
while minimizing noise from the incoherent portion of the coupling
(see Appendix~\ref{sec:quantum_classical}).

The classical equations %corresponding to 
for
the model in Eq.~\eqref{eq:master_equation} are
\vspace{-3mm}
\begin{subequations}
\label{eq:classical_model_coupled}
    \begin{align}
        \dot{\theta}_n =& \sin(\theta_n)\left[ \gamma_\text{g} - \gamma_\text{d} \sin^2(\theta_n) \right] + \sum_{m=1}^N \sin(\theta_m)\notag \\\times&\left[ A_{nm} \cos(\phi_m - \phi_n) - B_{nm} \sin(\phi_m - \phi_n) \right],
        \label{subeq:theta_eq_coupled}
    \end{align}
    \vspace{-0.8cm}
    \begin{align}
        \dot{\phi}_n =& \;\omega + \cot(\theta_n)\sum_{m=1}^N \sin(\theta_m)\notag\\\times&\left[ A_{nm}\sin(\phi_m - \phi_n) + B_{nm}\cos(\phi_m - \phi_n) \right],
        \label{subeq:phi_eq_coupled}
    \end{align}
\end{subequations}
where $\mathbf{C}=\mathbf{A}+{\rm i}\,\mathbf{B}$ for real matrices $\mathbf{A}$ and $\mathbf{B}$. In the absence of coupling, each spin asymptotically approaches and independently precesses around the stable limit cycle (assuming initial conditions in its attraction basin). Through the introduction of an appropriate choice of coupling in the form of $\mathbf{A}$ and $\mathbf{B}$, the spins can then converge to a common phase angle $\phi(t)$ and phase-synchronize. 
In general, the coupling modifies the angle $\theta_n$ of the limit cycles, as well as the precession frequencies, of the synchronized spins to values different from those in the uncoupled case.

\begin{figure*}
    \centering
    \includegraphics[width=\linewidth]{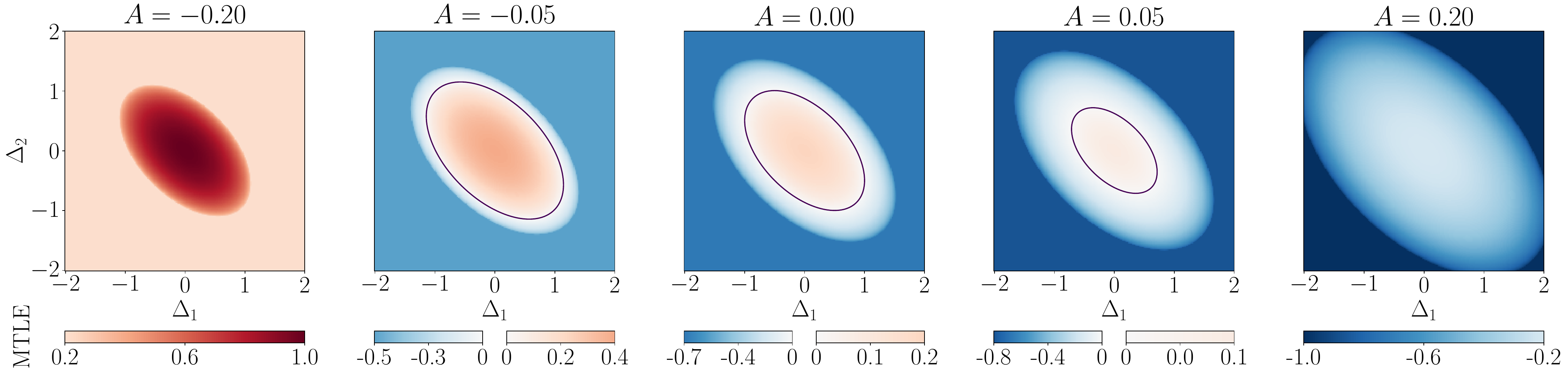}
    \caption{CSB in the classical model characterized by the MTLE in the $\Delta_1$--$\Delta_2$ plane. The complex coupling matrix is defined by 
    $A_{nm} = A$ for all $n$ and $m$, and  $B_{nm}$ given by Eq.~\eqref{eq:coupling}.  The MTLE becomes more negative as $A$ increases, indicating increased stability of the synchronized state in Eq.~\eqref{eq:sync_solution}. The black contour line separates the regions of positive and negative MTLE, with the middle three panels exhibiting CSB. The parameters are $\omega = 1$, $\gamma_\text{g} = 1$, $\gamma_\text{d} = 2$, and $c = 1$.}
    \label{fig:mtle}
\end{figure*}

\section{Synchronization in the classical model}
\label{sec:classical_sync}
The stability of the limit cycle at $\theta = \arcsin(\sqrt{\gamma_g/\gamma_d})$ in the single spin case suggests that $\theta$ is a natural candidate for synchronization upon the introduction of coupling between multiple spins. Indeed, for nonzero coupling matrices $\mathbf{A}$ and $\mathbf{B}$, one will find a synchronous solution of the system in Eq.~\eqref{eq:classical_model_coupled} given by
\begin{subequations}\label{eq:sync_solution}
    \begin{align}
        \theta^* &= \arcsin\left(\sqrt{\frac{\gamma_\text{g} + \sum_{k=1}^NA_{nk}}{\gamma_\text{d}}}\right), \label{subeq:sync_state_a}
        \\\phi^*(t) \! &=\!\! \left(\!\! \omega + \sqrt{1-\frac{\gamma_\text{g} + \sum_{k=1}^N A_{nk}}{\gamma_\text{d}}}\!\sum_{k=1}^N B_{nk} \!\! \right)\!\!t+\! \phi_0,
        \label{subeq:sync_state_b}
    \end{align}
\end{subequations}
provided that $\sum_{k=1}^N A_{nk}$ and $\sum_{k=1}^N B_{nk}$ are independent of $n$, where $\theta_n(t) = \theta^*$ and $\phi_n(t) = \phi^*(t)$ for $n=1, ...,N$, and that the quantities under the square roots are positive. It is convenient to write the synchronous solution in vector form
\begin{equation}
    \vec{X}(t) = (\theta_1(t), ..., \theta_N(t), \phi_1(t), ..., \phi_N(t)).\label{subeq:sync_vec}
\end{equation}
The stability of this solution can be assessed by evaluating the Jacobian of Eq.~\eqref{eq:classical_model_coupled} along it.

Setting $\dot{\theta}_n = f_n(\vec{\theta}, \vec{\phi})$ and $\dot{\phi}_n = g_n(\vec{\theta}, \vec{\phi})$, we can write the Jacobian in block form as
\begin{equation}\label{eq:jacobian}
    \renewcommand*{\arraystretch}{1.6}
    \mathbf{J}=\begin{pmatrix}
        \mathbf{M_{\theta}} & \mathbf{O_{\theta \phi}}\\
        \mathbf{O_{\phi \theta}} & \mathbf{M_{\phi}}
    \end{pmatrix},
\end{equation}
\noindent with the elements of each block matrix given by
\begin{subequations}
    \begin{align}
        (M_{\theta})_{nm} &= \left. \frac{\partial f_n}{\partial \theta_m} \right|_{\vec{X}} = \sqrt{1-\frac{\gamma_\text{g} + \sum_{k=1}^NA_{mk}}{\gamma_\text{d}}}\notag \\ & \;\;\;\;\;\;\;\;\times \left(A_{nm} - \delta_{nm}(2\gamma_\text{g} + 3\sum_{l=1}^NA_{ml}) \right),\\
        (O_{\theta \phi})_{nm} &= \left. \frac{\partial f_n}{\partial \phi_m} \right|_{\vec{X}} = \sqrt{\frac{\gamma_\text{g} + \sum_{k=1}^NA_{nk}}{\gamma_\text{d}}}\notag \\ &\;\;\;\;\;\;\;\;\;\;\;\;\;\times \left( - B_{nm} + \delta_{nm}\sum_{k=1}^N B_{nk} \right),\\
        (O_{\phi \theta})_{nm} &= \left. \frac{\partial g_n}{\partial \theta_m} \right|_{\vec{X}} = \sqrt{\frac{\gamma_\text{d}}{\gamma_\text{g} + \sum_{k=1}^NA_{nk}}}\notag \\&\!\!\!\!\!\!\!\!\!\!\!\!\!\!\!\!\!\!\!\!\!\!\!\times\left[ - \delta_{nm}\sum_{k=1}^N B_{nk} + \left( 1-\frac{\gamma_\text{g} + \sum_{k=1}^N A_{mk}}{\gamma_\text{d}} \right) B_{nm} \right],\\
        (M_{\phi})_{nm} &= \left. \frac{\partial g_n}{\partial \phi_m} \right|_{\vec{X}} = \sqrt{1 - \frac{\gamma_\text{g} + \sum_{k=1}^NA_{nk}}{\gamma_\text{d}}}\notag\\&\;\;\;\;\;\;\;\;\;\;\;\;\;\;\times\left( A_{nm} - \delta_{nm}\sum_{k=1}^NA_{mk} \right).
    \end{align}
\end{subequations}
The real parts of the Jacobian eigenvalues are the Lyapunov exponents. If all exponents are negative, except for one that is identically zero associated
with the direction along the trajectory, the solution $\vec{X}$ is stable with respect to perturbations. Stability is then determined by the largest exponent transverse to the trajectory, referred to as the maximum transverse Lyapunov exponent~(MTLE).

As a final note, just as in the single oscillator case, there is a second limit cycle at $\pi-\theta_n$. A synchronous solution with this adjusted angle and the same frequency as in Eq.~\eqref{subeq:sync_state_b} may be constructed. Evaluating the Jacobian, one will find a form identical to Eq.~\eqref{eq:jacobian} with the exception that the diagonal blocks are now negative, i.e., $\mathbf{M_{\theta}} \rightarrow -\mathbf{M_{\theta}}$ and $\mathbf{M_{\phi}} \rightarrow -\mathbf{M_{\phi}}$.

\section{Quantum converse symmetry breaking}

To establish our main results, we consider small $N$ both to limit the Hilbert-space dimension for large spin and because this regime already captures the essential effects, which can occur even more flexibly in larger networks.
Focusing on $N=3$ spins,
we explicitly introduce an adjacency matrix $\mathbf{C}$ of the form
\vspace{-2mm}
\begin{equation}\label{eq:coupling}
    \mathbf{C} = {\rm i}\,
    \begin{pmatrix}
    0 & -c/2 - \Delta_1 & c/2 + \Delta_1 \\
    c/2 + \Delta_2 & 0 & -c/2 - \Delta_2 \\
    -c/2 - \Delta_3 & c/2 + \Delta_3 & 0
    \end{pmatrix},
\end{equation} 
where $c$ characterizes the coupling strength and the $\Delta_n\text{'s}$ control the coupling heterogeneity, subject to the constraint $\sum_n\Delta_n=0$. When $\Delta_n=0$ for all $n$, the network has rotational node-permutation symmetry, whereas any other assignment of $\Delta_n$ breaks this symmetry. 
For real $c$ and $\Delta_n$ as considered here, the coupling matrix is purely imaginary (i.e., $\mathbf{C}={\rm i}\,\mathbf{B}$) and preserves cross-coupling between amplitude and phase dynamics.
This form of coupling leads to the possibility of CSB in the classical case, which persists for small nonzero $A_{nm}$.
Figure~\ref{fig:mtle} 
illustrates this persistence using the MTLE of the synchronous solution for
$\mathbf{C}=\mathbf{A} + {\rm i}\,\mathbf{B}$, where $A_{nm} = A \; \forall n,m$, with $A$ a constant, and $\mathbf{B}$ given by Eq.~\eqref{eq:coupling}. This robustness also suggests that the effect should extend over a finite range of noise in $\mathbf{C}$.

The synchronous solution for the coupling matrix in Eq.~\eqref{eq:coupling} is given by 
$\big(\theta_n(t), \phi_n(t)\big) =(\theta^*, \omega t +\phi_0)$ for all $n$ where $\theta^* = \arcsin(\sqrt{\gamma_\text{g}/\gamma_\text{d}})$ or $\pi - \arcsin(\sqrt{\gamma_\text{g}/\gamma_\text{d}})$, and $\phi_0$ is a constant phase common to all oscillators. One can verify that both synchronous solutions are unstable in the case of $\Delta_n$ equal to zero for each spin by calculating their MTLE from the Jacobian in Eq.~\eqref{eq:jacobian}. However, by making suitable choices in which the $\Delta_n$ are not all identical, we observe that the resulting coupling heterogeneity stabilizes a synchronous solution around the limit cycle at $\theta^* = \arcsin(\sqrt{\gamma_\text{g}/\gamma_\text{d}})$.

We can now turn to the quantum case.
In the absence of a quantum analog of the maximum transverse Lyapunov exponent, we introduce the spin variance operator:
$    
R = \frac{1}{N}\sum_{n=1}^N ||\vec{S}_n - \overline{\vec{S}}||^2,
$
where $\overline{\vec{S}} = \frac{1}{N}\sum_{n=1}^N \vec{S}_n$.
This operator has a minimum eigenvalue $\lambda_{\rm min}=S(N-1)/N$. An example, the case where the minimum eigenvalue is achieved is where all $N$ spins are in an identical coherent state, a special case being where they are all in the maximal eigenstate of the operator $S^z$.   
When $NS$ is an integer, as considered here, the operator has a maximum eigenvalue $\lambda_{\rm max}=S(S + 1)$ (see Supplemental Material (SM)~\cite{supp}, Sec.~\ref{sec:R_min_max}).
\nocite{atomic_coherent_states, glauber_spin_p, carmichael, Schwinger_rep, side_band_rev}
%SSS
%
To enable a fair comparison of the model across different values of $S$ and $N$, we define the normalized spin variance
\vspace{-1mm}
\begin{equation}\label{eq:rms}
    r = \frac{\langle R\rangle - \lambda_{\rm min}}{\lambda_{\rm max} - \lambda_{\rm min}}
\end{equation}
as a measure of synchronization, where $\langle \cdot \rangle$ is the quantum expectation value.

\begin{figure}
    \centering
    \includegraphics[scale=0.3]{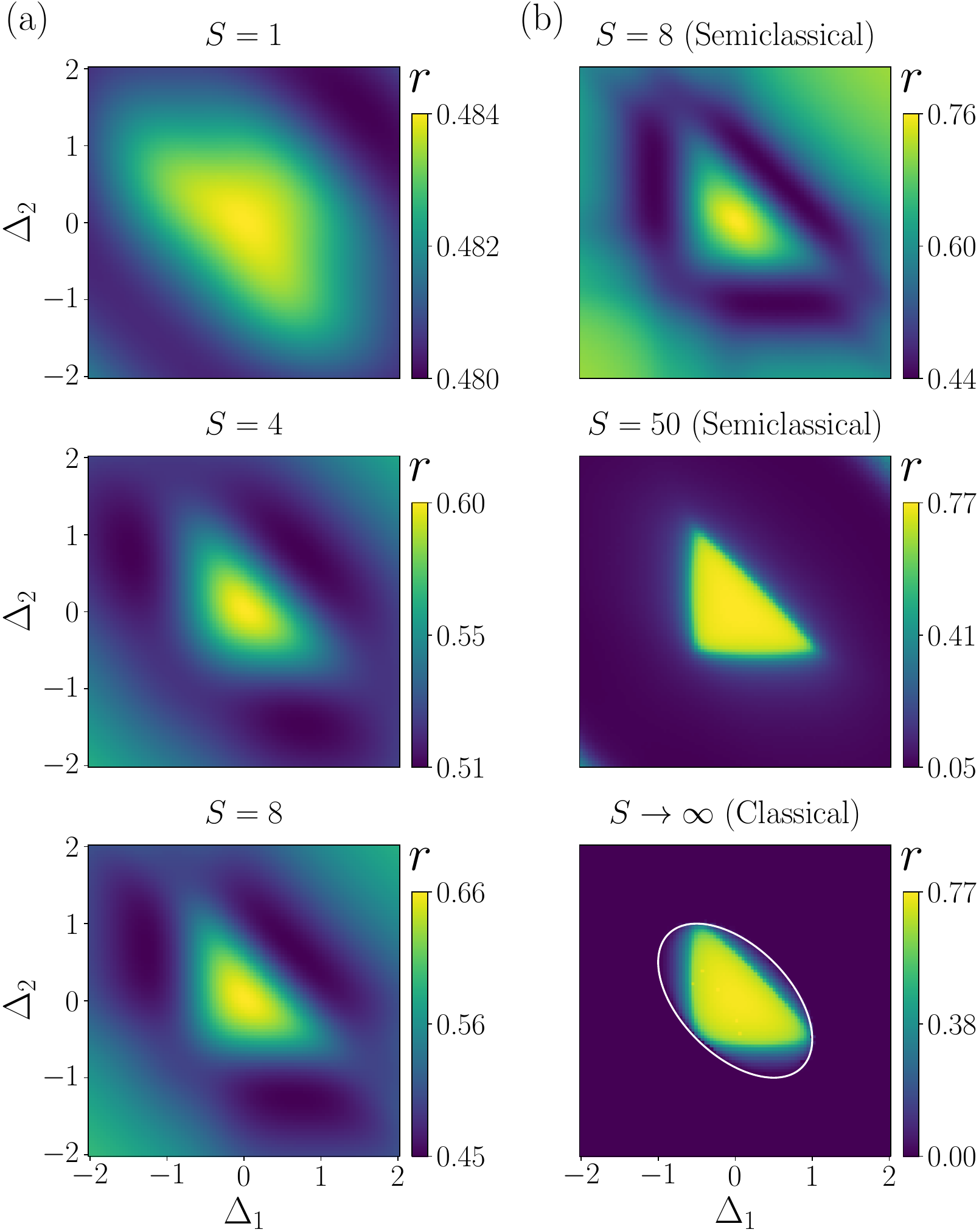}
    \vspace{-1mm}
    \caption{Quantum network dynamics and their transition to the classical regime in the semiclassical model. 
    Normalized spin variance $r$ as a function of coupling heterogeneity $\Delta_1$ and $\Delta_2$ (with $\Delta_3 = -\Delta_1 - \Delta_2$), shown for increasing spin $S$ in 
    (a)~quantum and (b)~semiclassical calculations. The quantum case is evaluated using the steady state of Eq.~\eqref{eq:master_equation}. The semiclassical case is obtained by time-averaging Eq.~\eqref{eq:langevin_eq} for $\epsilon= 0.5$, which, in the limit of large $S$ and in the absence of regularization, is equivalent to Eq.~\eqref{eq:classical_model_coupled}. In the classical case in (b), the white contour line encloses the points with $r>0$, separating the synchronous and asynchronous regions.
    In all panels, we use $\omega = 1$, $\gamma_\text{g} = 1$, $\gamma_\text{d} = 2$, and $c = 1$, as in Fig.~\ref{fig:mtle}. For details on the convergence timescales, see SM~\cite{supp}, Sec.~\ref{sec:timescale}.
    %SSS
   }
    \label{fig:RMS}
\end{figure}

Figure~\ref{fig:RMS}(a) 
shows $r$ as a function of coupling heterogeneity when the expectation value in Eq.~\eqref{eq:rms} is evaluated in the steady state  $\rho_\text{ss}$,
which is defined as the solution of the master equation Eq.~\eqref{eq:master_equation} with $\dot{\rho}=0$.
For different spin values $S$, 
this quantity is consistently largest, and the system is thus least synchronized, when all spins are identically coupled
($\Delta_{1,2} = 0$), and it decreases with suitable heterogeneity. This can be interpreted as a quantum manifestation of CSB, in the sense that a more synchronous (and thus more symmetric) state is promoted by symmetry-breaking heterogeneity. The range of $r$ induced by heterogeneity increases with spin,
indicating that the effect becomes more pronounced as $S$ increases. These results are significant for two reasons. First, because CSB is not exhibited by all classical oscillator networks, it was not known whether it would emerge in the classical limit of the coupled quantum systems. Second, the notion of stabilizing synchronous orbits has no direct counterpart in the quantum regime, making the observation of an analogous quantum effect unexpected a priori.

Due to the rapid growth of the Hilbert space dimension with increasing spin, simulations of the exact quantum model quickly become intractable. For this reason, we introduce a  Langevin equation for the variable $u_n = \tan(\theta_n/2)e^{{\rm i}\,\phi_n}$, which approximates the statistics of the exact model in Eq.~\eqref{eq:master_equation} for larger spins, as follows:
{\small
\begin{subequations}\label{eq:langevin_eq}
\begin{equation}\label{subeq:langevin_eq}
        \dot{u}_n = \mu_n + \sigma_n \big(\xi_n^\text{R}(t) + {\rm i}\,\xi_n^\text{I}(t)\big) + M(u_n),
    \end{equation}
    \vspace{-0.6cm}
    \begin{align}\label{subeq:drift}
        &\mu_n = {\rm i}\, \omega u_n + \!\Big[ \gamma_\text{g} (S\!+\!1)\!+ \!\!\sum_{m=1}^N\!C_{nm}^{h+}\Big]\frac{u_n}{S}
        - \gamma_\text{d} \frac{2S\!-\!1}{S^2}\notag\\&\times\!\frac{2S + |u_n|^2}{(1 + |u_n|^2)^2} \left|u_n\right|^2u_n + \!\sum_{m=1}^N \!\frac{C_{nm}u_m + C_{nm}^* u_n^2 u_m^*}{1 + |u_m|^2},
    \end{align}
        \vspace{-0.4cm}
\begin{align}\label{subeq:diffusion}
        \sigma_n = \sqrt{ \frac{\gamma_\text{g}}{2S} + \frac{\gamma_\text{d}}{2S^2} |u_n|^6\left(\frac{4(S-1) - 2|u_n|^2}{(1+|u_n|^2)^2} + \frac{1}{S}\right)},
    \end{align}
        \vspace{-0.1cm}
\begin{equation}\label{subeq:regularization}
        M(u_n) = \left(1-e^{\epsilon|u_n|^2}\right)\frac{u_n}{|u_n|},
    \end{equation}
\end{subequations}}
where $\xi_n^\text{R}(t)$ and $\xi_n^\text{I}(t)$ are independent Gaussian white noise processes with zero mean, such that $\langle \xi_n^\text{R}(t) \xi_m^\text{R}(t') \rangle = \langle \xi_n^\text{I}(t) \xi_m^\text{I}(t') \rangle = \delta_{nm}\delta(t-t')$ and $\langle \xi_n^\text{R}(t) \xi_m^\text{I}(t') \rangle = 0$ $\forall\, n, m$.
(With a slight abuse of notation, $\langle \cdot \rangle$ now denotes a classical expectation value.)
This equation was derived by first finding the Glauber-Sudarshan $P$-distribution equation corresponding to Eq.~\eqref{eq:master_equation} and then dropping the derivatives of order higher than two, any second-order terms that could compromise the positive semidefiniteness of the diffusion matrix, and, for computational convenience, any noise terms proportional to the coupling (see SM~\cite{supp}, Sec.~\ref{sec:p_distribution}).
%SSS
The resulting Fokker–Planck equation specifies the drift term in Eq.~\eqref{subeq:drift} and the diffusion term in Eq.~\eqref{subeq:diffusion} that define the corresponding Langevin equation \eqref{subeq:langevin_eq}.
The term in Eq.~\eqref{subeq:regularization} is a regularization term not derived from the original model, whose magnitude is controlled by the parameter $\epsilon$. This term prevents the noise from driving the spins into the basin of attraction of the fixed point at infinity ($|u_n|\rightarrow\infty$), which causes difficulties when integrating numerically. The solution
$u_n$ of Eq.~\eqref{subeq:langevin_eq} parametrizes the spin coherent state $|u_n(\theta_n, \phi_n)\rangle = e^{-{\rm i}\,\phi_n S^z}e^{-{\rm i}\,\theta_n S^y}|S,S\rangle$, where $|S,S\rangle$ is the maximal eigenstate of $S^z$ with spin quantum number $S$.

To calculate the steady-state expectation of $R$ in this semiclassical formulation,
we take the time-average of $\langle\vec{u}(t)|R|\vec{u}(t)\rangle$ over a sufficiently long time period, where $|\vec{u}(t)\rangle = \bigotimes_{n=1}^N|u_n(t)\rangle$.
The resulting $r$ is plotted for representative values of $S$ in Fig.~\ref{fig:RMS}(b); here,
in addition to time averaging, averaging over multiple trajectories originating from independent initial conditions $\vec{u}(0)$ was performed. Simulations show that the semiclassical and quantum cases share the same qualitative features, as illustrated for $S=8$ by the top panel of Fig.~\ref{fig:RMS}(b) and the bottom panel of Fig.~\ref{fig:RMS}(a), respectively. In the limit of large $S$ and small $\epsilon$ (no regularization),  Eq.~\eqref{subeq:langevin_eq} converges to the classical model in Eq.~\eqref{eq:classical_model_coupled}.
While the bottom panel of Fig.~\ref{fig:RMS}(b) does not strictly correspond to the purely classical model in Eq.~\eqref{eq:classical_model_coupled}, since the regularization parameter $\varepsilon$ remains finite, it can be regarded as effectively classical because the majority of the points on the sphere are mapped to a relatively small disk in the complex plane containing the $u_n$. The white contour in this panel encloses the $r>0$ region, which closely matches the region of positive MTLE in Fig.~\ref{fig:mtle} for $A=0$.

\begin{figure}
    \centering
    \includegraphics[scale=0.21]{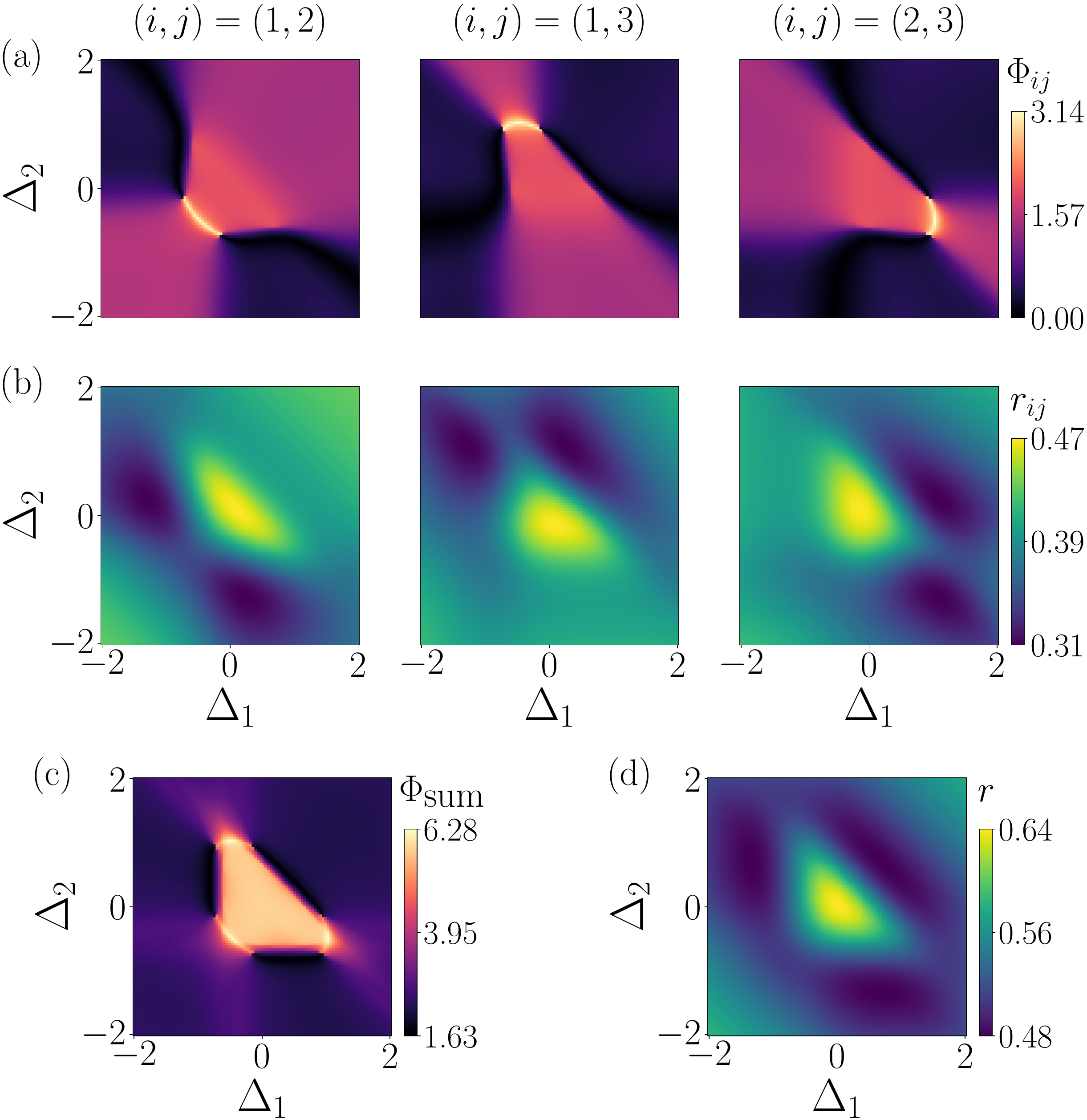}
    \vspace{-1mm}
    \caption{
    Pairwise quantum dynamics in the 3-spin network for $S=6$, according to two synchronization measures. Pairwise (a) effective phase-locking angle $\Phi_{ij}$
    and (b) normalized spin variance $r_{ij}$, arranged from left to right as $(i,j)=$ (1, 2), (1, 3), and (2, 3).  (c) Pixelwise sum of the pairwise $\Phi_{ij}$ in (a). (d) Normalized spin variance over all oscillators obtained from Eq.~\eqref{eq:rms} or, equivalently, from a rescaled sum of the pairwise $r_{ij}$ in (b) (see SM~\cite{supp}, Sec.~\ref{sec:pairwise_r}).
    %SSS 
    The other parameters are the same as in Fig.~\ref{fig:RMS}.
    }
    \label{fig:RMS_vs_phase}
\end{figure}
 
\section{Pairwise quantum synchronization}

Phase locking between spin pairs has been previously studied \cite{relphase, spin1_entanglement} and can be viewed as a herald of synchronization \cite{sync}.
The effective phase
angle between two spins can be defined as
\begin{equation}\label{eq:rel_phase}
    \Phi_{ij} = \Big|\underset{-\pi<\phi\leq\pi}{\text{argmax}}\big(S_{ij}(\phi)\big)\Big|,
\end{equation}
where $ S_{ij}(\phi)\! =\! - \frac{1}{2\pi} +\! \iiint d\phi_j\, d\theta_i\, d\theta_j \sin(\theta_i)\sin(\theta_j)Q(\cdot)$
is the relative phase distribution between spins $i$ and $j$ ~\cite{relphase}.
In this formula,
$Q(\cdot)=Q(\theta_i, \theta_j, \phi_i, \phi_j)$
is the Husimi $Q$-function of the reduced density matrix with all spins but $i$ and $j$ traced out, and the relative phase
$\phi$ is used in the change of variables $\phi_i=\phi+\phi_j$ under the integral. Note that this definition of the phase locking angle is only sensible if $S_{ij}(\phi)$ is unimodal, which is dominantly the case for the range of ($\Delta_1$, $\Delta_2$) considered.
Figure~\ref{fig:RMS_vs_phase}(a) shows the result of evaluating Eq.~\eqref{eq:rel_phase} for each spin pair in the case of $S=6$.
The region of low $\Phi_{ij}$ rotates by approximately $120^\circ$ from one plot to the next across the row.
The limited overlap among these regions indicates that, for most coupling-heterogeneity assignments, at most one spin pair is phase locked.

To enable comparison with the full network analysis, we introduce a pairwise normalized spin variance $r_{ij}$, defined analogously to Eq.~\eqref{eq:rms} but evaluated for $N=2$, with the sum restricted to the selected nodes $i$ and $j$. As shown in Fig.~\ref{fig:RMS_vs_phase}(b), the regions of low variance exhibit the same three-fold rotational symmetry observed for the regions of zero relative phase in Fig.~\ref{fig:RMS_vs_phase}(a). 
Again, comparison across panels shows a lack of concurrent strong synchronization among all spins in the ($\Delta_1$, $\Delta_2$) plane.

Adding the pairwise contributions together and applying the appropriate rescaling yields the total normalized variance presented in Fig.~\ref{fig:RMS_vs_phase}(d),
which is the quantity used in Fig.~\ref{fig:RMS}. 
As shown in Fig.~\ref{fig:RMS_vs_phase}(c),
this agrees well
with the
result of summing over the pairwise phase locking angles in Fig.~\ref{fig:RMS_vs_phase}(a).
The consistency between the two synchronization measures, $\Phi_{ij}$ and $r_{ij}$, is further confirmed by comparing them for individual spin pairs.
These results indicate the occurrence of CSB in the reduced pairwise systems, as the synchronization regions for a given pair $(i,j)$ consistently occur for distinct values of $(\Delta_i, \Delta_j)$.
This contrasts with the corresponding classical case, where heterogeneity can promote 
synchronization among all oscillators (see SM~\cite{supp}, Fig.~\ref{fig:classical_pairwise}),
%SSS 
even though  $r_{ij}$ and $\Phi_{ij}$ converge to their classical values as $S\rightarrow\infty$ 
and the noise in the semiclassical model vanishes.
In all cases, the equilibrium values of $\Phi_{ij}$ and $r_{ij}$ (and thus of $r$) are not imposed but emerge from the system dynamics. Accordingly, CSB is best characterized by the changes in $r_{ij}$ and $\Phi_{ij}$ rather than by their absolute values.

The coupled spin vdP oscillators considered here can be realized with trapped ions using the Schwinger representation, with dissipation implemented via sideband cooling (see SM \cite{supp}, Sec.~\ref{sec:experiment}).
%SSS
The measures $r_{ij}$ and $\Phi_{ij}$ are then obtained from the reconstructed density matrix using quantum state tomography~\cite{gaussian_tomography}.

\section{Conclusion}

Our demonstration that heterogeneous coupling outperforms homogeneous coupling in synchronizing spin van der Pol oscillators extends the phenomenon of converse symmetry breaking to the quantum regime. This reveals important connections and distinctions between classical and quantum synchronization. 
Heterogeneity in quantum synchronization was previously emphasized in quantum synchronization blockade \cite{qsync_blockade}, where identical oscillators fail to synchronize, whereas suitably detuned ones synchronize due to energy quantization and quantum selection rules; consequently, synchronization blockade disappears in the classical regime. 
The mechanism identified here has a fundamentally different origin and is enabled by a symmetry phenomenon already present in the classical limit \cite{AEMprl2016}, where synchronous states with a given symmetry are stabilized by explicitly breaking the corresponding system symmetry through suitable heterogeneity.
The present effect is therefore reminiscent of the induced stability and is expected to be common in quantum systems whose classical limits exhibit converse symmetry breaking,  including bosonic quantum vdP oscillators \cite{qsync_expt2}.

At the same time, we identify clear quantum signatures,
notably the tendency for synchronization in the small-spin regime to occur predominantly 
between spin pairs rather than across the full system.
Given known relations between synchronization and entanglement \cite{spin1_entanglement, dirk2017}, coupling mismatches may provide a useful resource for entanglement generation in quantum networks.  This work can be extended by considering larger networks which may reveal 
topological features of quantum synchronization \cite{topological_qsync}.

\let\oldaddcontentsline\addcontentsline% Store \addcontentsline
\renewcommand{\addcontentsline}[3]{}% Make \addcontentsline a no-op

%\vspace{1mm}

\section*{Acknowledgments}

The authors acknowledge support from ARO MURI Grant No.~W911NF-21-1-0325.

\section*{Data availability}
The code used to generate the data in this study is openly available~\cite{github_spincsb}.\nocite{qutip}

{\appendix

\section{Relationship between quantum and classical models}
\label{sec:quantum_classical}

The general form of the Lindblad master equation for $N$ coupled spins is
\begin{equation}\label{eq:quantum_model}
    \dot{\rho} = -{\rm i}\,[H, \rho] + \mathcal{L}_{\text{ds}}\rho + \mathcal{L}_{\text{cp}}\rho,
\end{equation}
where
\begin{subequations}\label{eq:q_model_terms}
    \begin{align}
        H &= \sum_{n=1}^N \omega S_n^z, \label{subeq:hamiltonian}\\
        \mathcal{L}_{\text{ds}}(\cdot) &= \sum_{n=1}^N \Big(\gamma_\text{g} \mathcal{D}[{S'_n}^-](\cdot) + \frac{\gamma_\text{d}}{2 S}\mathcal{D}[({S'_n}^+)^2](\cdot)\Big), \label{subeq:dissipator}\\
        \mathcal{L}_{\text{cp}}(\cdot) &= \sum_{n=1}^N \sum_{m=1}^N \bigg(\frac{1}{2}C_{nm}^a[{S'_n}^{-} {S'_m}^+, \cdot] \notag\\
        &\!\!\!\!\!\!\!\!\!\!\!\!\!\!+ C_{nm}^{h+}\mathcal{D}[{S'_n}^-, {S'_m}^+](\cdot) - C_{nm}^{h-^*}\mathcal{D}[{S'_n}^+, {S'_m}^-](\cdot)\bigg),
        \label{subeq:dissipator_coupling}
    \end{align}
\end{subequations}
and ${S'_n}^{\pm} = S_n^{\pm}/\sqrt{S}$ are the reduced ladder operators. The adjacency matrix elements $C_{nm}$ of the spin network are the entries of a general $N\times N$ complex valued matrix, with each $C_{nm}$ equal to the sum of $C_{nm}^a$, $C_{nm}^{h+}$, and  $C_{nm}^{h-}$. To obtain these matrix elements from $C_{nm}$, we first decompose $C_{nm}$ into its Hermitian and anti-Hermitian parts: $\mathbf{C}^h = \frac{1}{2}\left(\mathbf{C}+\mathbf{C}^\dag \right)$ and $\mathbf{C}^a = \frac{1}{2}\left(\mathbf{C}-\mathbf{C}^\dag \right)$. We then diagonalize $\mathbf{C}^h$ into $\mathbf{C}^h = UDU^\dag = UD^+U^\dag + UD^-U^\dag$, where $D$ and $U$ are the diagonal eigenvalue and unitary diagonalizing matrix, respectively. We set $\mathbf{C}^{h\pm} = UD^\pm U^\dag$, where $D^+$ ($D^-$) contains the entries of $D$ larger (smaller) than zero, so that $D = D^+ + D^-$. This decomposition of the Hermitian part of $\mathbf{C}$ into $\mathbf{C}^{h+}$ and $\mathbf{C}^{h-}$ is not unique, as one can always add a positive semidefinite matrix to $\mathbf{C}^{h+}$ and subtract the same from $\mathbf{C}^{h-}$. However, it is the unique ``minimal" decomposition in the sense that there is no positive semidefinite matrix that can be subtracted from $\mathbf{C}^{h+}$ and added to $\mathbf{C}^{h-}$ that can give another valid decomposition. This choice minimizes the noise introduced by the incoherent coupling terms and can be most clearly seen by the presence of $\mathbf{C}^{h\pm}$ in the diffusion term of the equation of motion for the $P$ phase-space distribution in SM~\cite{supp}, Sec.~\ref{sec:p_distribution}.

From these equations, we can derive the corresponding classical equations by assigning the classical variable $\vec{L}_n = \langle \vec{S}_n \rangle$. Performing this while keeping only terms at leading order in $\hbar$ and approximating $\langle O_1O_2\rangle \approx \langle O_1 \rangle \langle O_2 \rangle$ for operators $O_1$ and $O_2$, one obtains
\begin{align}
    &\dot{\vec{L}}_n = \vec{L}_n \times \left[-\omega\hat{z} + \vec{L}_n \times \hat{z} \Big[ \frac{\gamma_\text{g}}{S} - \frac{\gamma_\text{d}}{S^3}\Big( S^2 - \big( \vec{L}_n \cdot \hat{z} \big)^2 \Big) \Big] \right]\notag \\ 
    &\!\!\!+ \frac{\vec{L}_n}{S} \times \Big[ \sum_{m=1}^N A_{nm}\vec{L}_m \times \hat{z} \Big] + \frac{\vec{L}_n}{S} \times\Big[\hat{z}\times\sum_{m=1}^N B_{nm}\vec{L}_m \times \hat{z} \Big],
\end{align}
 where $\mathbf{C} = \mathbf{A} + {\rm i}\,\mathbf{B}$ for real matrices $\mathbf{A}$ and $\mathbf{B}$. Note that $\dot{\vec{L}}_n \cdot \vec{L}_n = 0$, meaning that the norm of each spin vector is conserved. This should not be surprising since the spin quantum number $S$ of each oscillator is fixed in the quantum model. The classical model in Eq.~\eqref{eq:classical_model_coupled}
 is then obtained by performing a change of variables to angular coordinates $\theta_n$ and $\phi_n$. From the angular form of the classical equations, it is clear that the classical system will exhibit fixed-amplitude oscillations similar to the Stuart-Landau model. In the absence of coupling ($\mathbf{A} = \mathbf{B} = 0$), setting $\dot{\theta}_n$ equal to zero one obtains four possible solutions for $\theta_n$. Two of these solutions correspond to fixed points at $\theta_n=0$ and $\pi$, while the other two are limit cycles with amplitudes $\theta_n = \arcsin(\sqrt{\gamma_\text{g}/\gamma_\text{d}})$ and $\pi - \arcsin(\sqrt{\gamma_\text{g}/\gamma_\text{d}})$. Evaluating the derivative of $\dot{\theta}_n$ with respect to $\theta_n$ at the limit cycle amplitudes, one will find the former to be stable while the latter is unstable.
 }

\bibliography{apssamp}% Produces the bibliography via BibTeX.

@book{sync,
    author = "A. S. Pikovsky and M. Rosenblum and J. Kurths",
    title = "\uppercase{S}ynchronization: \uppercase{A} \uppercase{U}niversal \uppercase{C}oncept in \uppercase{N}onlinear \uppercase{S}ciences",
    publisher = "Cambridge University Press", address = "New York",
    year = "2001"
}

@incollection{open_systems,
    author = {Breuer, Heinz-Peter and Petruccione, Francesco},
    isbn = {9780199213900},
    title = {\uppercase{Q}uantum \uppercase{M}aster \uppercase{E}quations},
    booktitle = {\uppercase{T}he \uppercase{T}heory of \uppercase{O}pen \uppercase{Q}uantum \uppercase{S}ystems},
    publisher = {Oxford University Press},
    address   = {Oxford},
    year = {2007},
    month = {01},
    doi = {10.1093/acprof:oso/9780199213900.003.03}
}

@article{kato2025,
  title = {Quantum spin van der {P}ol oscillator: Spin-based limit-cycle oscillator exhibiting quantum synchronization},
  author = {Kato, Yuzuru and Nakao, Hiroya},
  journal = {Phys. Rev. Res.},
  volume = {7},
  issue = {4},
  pages = {043133},
  numpages = {13},
  year = {2025},
  month = {Nov},
  publisher = {American Physical Society},
  doi = {10.1103/mfvy-bbzm},
  url = {https://link.aps.org/doi/10.1103/mfvy-bbzm}
}

@ARTICLE{damped_spin,
  author={Mallinson, J.},
  journal={IEEE Trans. Magn.}, 
  title={On damped gyromagnetic precession}, 
  year={1987},
  volume={23},
  number={4},
  pages={2003-2004},
  doi={10.1109/TMAG.1987.1065181}}

@article{Lee_prl,
   author       = "T. E. Lee and H. R. Sadeghpour",
   title        = "Quantum Synchronization of Quantum van der {P}ol Oscillators with Trapped Ions",
   journal      = "Phys. Rev. Lett.",
   volume       = "111",
   pages        = "234101", 
   year         = "2013"
}

@article{Bruder_prl1,
   author       = "S. Walter and A. Nunnenkamp and C. Bruder", 
   title        = "Quantum Synchronization of a Driven Self-Sustained Oscillator", 
   journal      = "Phys. Rev. Lett.", 
   volume       = "112", 
   pages        = "094102", 
   year         = "2014"
}

@article{lee_pre,
  title = {Entanglement tongue and quantum synchronization of disordered oscillators},
  author = {Tony E. Lee and Ching-Kit Chan and Shenshen Wang},
  journal = {Phys. Rev. E},
  volume = {89},
  pages = {022913},
  year = {2014}
}

@article{Bruder_annalender,
   author       = "S. Walter and A. Nunnenkamp and C. Bruder",
   title        = "Quantum synchronization of two van der {P}ol oscillators",
   journal      = "Ann. Phys. (Berlin)",
   volume       = "527",
   pages        = "131-138",
   year         = "2015"
}

@article{relphase,
   author       = "C. Davis-Tilley and A. D. Armour",
   title        = "Synchronization of micromasers",
   journal      = "Phys. Rev. A",
   volume       = "94",
   pages        = "063819",
   year         = "2016"
}

@article{AEMprl2016,
  title = {Symmetric States Requiring System Asymmetry},
  author = {Nishikawa, Takashi and Motter, Adilson E.},
  journal = {Phys. Rev. Lett.},
  volume = {117},
  issue = {11},
  pages = {114101},
  numpages = {5},
  year = {2016},
  month = {Sep},
  publisher = {American Physical Society},
  doi = {10.1103/PhysRevLett.117.114101},
  url = {https://link.aps.org/doi/10.1103/PhysRevLett.117.114101}
}

@article{qsync_blockade,
  title = {Quantum Synchronization Blockade: Energy Quantization Hinders Synchronization of Identical Oscillators},
  author = {L\"orch, Niels and Nigg, Simon E. and Nunnenkamp, Andreas and Tiwari, Rakesh P. and Bruder, Christoph},
  journal = {Phys. Rev. Lett.},
  volume = {118},
  issue = {24},
  pages = {243602},
  numpages = {6},
  year = {2017},
  month = {Jun},
  publisher = {American Physical Society},
  doi = {10.1103/PhysRevLett.118.243602},
  url = {https://link.aps.org/doi/10.1103/PhysRevLett.118.243602}
}

@article{blockade_nigg,
  title = {Observing quantum synchronization blockade in circuit quantum electrodynamics},
  author = {Nigg, Simon E.},
  journal = {Phys. Rev. A},
  volume = {97},
  issue = {1},
  pages = {013811},
  numpages = {10},
  year = {2018},
  month = {Jan},
  publisher = {American Physical Society},
  doi = {10.1103/PhysRevA.97.013811},
  url = {https://link.aps.org/doi/10.1103/PhysRevA.97.013811}
}

@article{interference_blockade,
  title = {Quantum synchronization through the interference blockade},
  author = {Kehrer, Tobias and Nadolny, Tobias and Bruder, Christoph},
  journal = {Phys. Rev. A},
  volume = {110},
  issue = {4},
  pages = {042203},
  numpages = {10},
  year = {2024}
}

@article{nonreciprocal_blockade,
  title = {Quantum synchronization blockade induced by nonreciprocal coupling},
  author = {Kehrer, Tobias and Bruder, Christoph},
  journal = {Phys. Rev. A},
  volume = {112},
  pages = {012223},
  numpages = {12},
  year = {2025}
}

@article{measurement_Kato,
doi = {10.1088/1367-2630/abd7be},
url = {https://dx.doi.org/10.1088/1367-2630/abd7be},
year = {2021},
month = {jan},
publisher = {IOP Publishing},
volume = {23},
number = {1},
pages = {013007},
author = {Yuzuru Kato and Hiroya Nakao},
title = {Enhancement of quantum synchronization via continuous measurement and feedback control},
journal = {New J. Phys.}
}

@article{measurement_kwek,
  title = {Enhancing quantum synchronization through homodyne measurement, noise, and squeezing},
  author = {Shen, Yuan and Soh, Hong Yi and Fan, Weijun and Kwek, L. C.},
  journal = {Phys. Rev. E},
  volume = {108},
  issue = {2},
  pages = {024204},
  numpages = {7},
  year = {2023},
  month = {Aug},
  publisher = {American Physical Society},
  doi = {10.1103/PhysRevE.108.024204},
  url = {https://link.aps.org/doi/10.1103/PhysRevE.108.024204}
}

@article{measurement_bruder,
  title = {Quantum limit cycles and synchronization from a measurement perspective},
  author = {Nadolny, Tobias and Bruder, Christoph},
  journal = {Phys. Rev. Res.},
  volume = {8},
  issue = {2},
  pages = {023050},
  numpages = {14},
  year = {2026},
  month = {Apr},
  publisher = {American Physical Society},
  doi = {10.1103/c5wj-4s73},
  url = {https://link.aps.org/doi/10.1103/c5wj-4s73}
}

@article{squeezing,
  title = {Squeezing Enhances Quantum Synchronization},
  author = {Sonar, Sameer and Hajdu\ifmmode \check{s}\else \v{s}\fi{}ek, Michal and Mukherjee, Manas and Fazio, Rosario and Vedral, Vlatko and Vinjanampathy, Sai and Kwek, Leong-Chuan},
  journal = {Phys. Rev. Lett.},
  volume = {120},
  issue = {16},
  pages = {163601},
  numpages = {5},
  year = {2018},
  month = {Apr},
  publisher = {American Physical Society},
  doi = {10.1103/PhysRevLett.120.163601},
  url = {https://link.aps.org/doi/10.1103/PhysRevLett.120.163601}
}

@article{quadratic_qsync,
  title = {Quantum synchronization in quadratically coupled quantum van der {P}ol oscillators},
  author = {Nissi Thomas and M. Senthilvelan},
  journal = {Phys. Rev. A},
  volume = {106},
  pages = {012422},
  year = {2022}
}

@article{noise_qsync_2016,
  title = {Noise-induced transitions in optomechanical synchronization},
  author = {Talitha Weiss and Andreas Kronwald and Florian Marquardt},
  journal = {New J. Phys.},
  volume = {18},
  pages = {013043},
  year = {2016}
}

@article{noise_qsync_2022,
  title = {Noise-Induced Quantum Synchronization},
  author = {Schmolke, Finn and Lutz, Eric},
  journal = {Phys. Rev. Lett.},
  volume = {129},
  issue = {25},
  pages = {250601},
  numpages = {7},
  year = {2022}
}

@article{noise_qsync_expt,
  title = {Noise-induced quantum synchronization with entangled oscillations},
  author = {Tao, Ziyu and Schmolke, Finn and Hu, Chang-Kang and Huang, Wenhui and Zhou, Yuxuan and Zhang, Jiawei and Chu, Ji and Zhang, Libo and Sun, Xuandong and Guo, Zechen and Niu, Jingjing and Weng, Wenle and Liu, Song and Zhong, Youpeng and Tan, Dian and Yu, Dapeng and Lutz, Eric},
  journal = {Nat. Commun.},
  volume = {16},
  pages = {8457},
  year = {2025}
}

@article{nonlinearity_qsync,
  title = {Quantum synchronization effects induced by strong nonlinearities},
  author = {Yuan Shen and W. Mok and Changsuk Noh and Ai Qun Liu and L. C. Kwek and Weijun Fan and Andy Chia},
  journal = {Phys. Rev. A},
  volume = {107},
  pages = {053713},
  year = {2023}
}

@article{relaxation_qsync,
  title = {Relaxation oscillations and frequency entrainment in quantum mechanics},
  author = {A. Chia and L. C. Kwek and C. Noh},
  journal = {Phys. Rev. E},
  volume = {102},
  pages = {042213},
  year = {2020}
}

@article{topological_qsync,
  title = {Topological synchronization of quantum van der {P}ol oscillators},
  author = {Christopher W. Wächtler and Gloria Platero},
  journal = {Phys. Rev. Res.},
  volume = {5},
  pages = {023021},
  year = {2023}
}

@article{qsync_expt1,
author = {Yi Li  and Zihan Xie  and Xiaodong Yang  and Yue Li  and Xingyu Zhao  and Xu Cheng  and Xinhua Peng  and Jun Li  and Eric Lutz  and Yiheng Lin  and Jiangfeng Du },
title = {Experimental realization and synchronization of a quantum van der {P}ol oscillator},
journal = {Sci. Adv.},
volume = {11},
number = {41},
pages = {eady5649},
year = {2025}
}

@article{qsync_expt2,
  title = {Observation of Synchronization between Two Quantum van der {P}ol Oscillators in Trapped Ions},
  author = {Liu, Jiarui and Wu, Qiming and Moore, Joel E. and Haeffner, Hartmut and W\"achtler, Christopher W.},
  journal = {Phys. Rev. X},
  volume = {16},
  issue = {2},
  pages = {021062},
  numpages = {19},
  year = {2026},
  month = {Jun},
  publisher = {American Physical Society},
  doi = {10.1103/w1bm-wjl4},
  url = {https://link.aps.org/doi/10.1103/w1bm-wjl4}
}

@article{BruderSmall,
  title = {Synchronizing the Smallest Possible System},
  author = {Roulet, Alexandre and Bruder, Christoph},
  journal = {Phys. Rev. Lett.},
  volume = {121},
  issue = {5},
  pages = {053601},
  numpages = {5},
  year = {2018},
  month = {Jul},
  publisher = {American Physical Society},
  doi = {10.1103/PhysRevLett.121.053601},
  url = {https://link.aps.org/doi/10.1103/PhysRevLett.121.053601}
}

@article{spin1_entanglement,
  title = {Quantum Synchronization and Entanglement Generation},
  author = {Roulet, Alexandre and Bruder, Christoph},
  journal = {Phys. Rev. Lett.},
  volume = {121},
  issue = {6},
  pages = {063601},
  numpages = {5},
  year = {2018},
  month = {Aug},
  publisher = {American Physical Society},
  doi = {10.1103/PhysRevLett.121.063601},
  url = {https://link.aps.org/doi/10.1103/PhysRevLett.121.063601}
}

@article{optimal_spin_qsync,
  title = {Optimal synchronization deep in the quantum regime: Resource and fundamental limit},
  author = {Martin Koppenh\"ofer and Alexandre Roulet},
  journal = {Phys. Rev. A},
  volume = {99},
  pages = {043804},
  year = {2019}
}

@article{macroscopic_qsync,
  title = {Macroscopic Quantum Synchronization Effects},
  author = {Tobias Nadolny and Christoph Bruder},
  journal = {Phys. Rev. Lett.},
  volume = {131},
  pages = {190402},
  year = {2023}
}

@article{spin1_expt,
  title = {Observation of Quantum Phase Synchronization in Spin-1 Atoms},
  author = {Arif Warsi Laskar and Pratik Adhikary and Suprodip Mondal and Parag Katiyar and Sai Vinjanampathy and Saikat Ghosh},
  journal = {Phys. Rev. Lett.},
  volume = {125},
  pages = {013601},
  year = {2020}
}

@article{
    Zhang2021,
    author = {Yuanzhao Zhang  and Jorge L. Ocampo-Espindola  and István Z. Kiss  and Adilson E. Motter },
    title = {Random heterogeneity outperforms design in network synchronization},
    journal = {Proc. Natl. Acad. Sci. USA},
    volume = {118},
    number = {21},
    pages = {e2024299118},
    year = {2021},
    doi = {10.1073/Proc. Natl. Aca. Sci. USA.2024299118}
}

@article{trapped_ions,
  title = {Realization of a Quantum Integer-Spin Chain with Controllable Interactions},
  author = {Senko, C. and Richerme, P. and Smith, J. and Lee, A. and Cohen, I. and Retzker, A. and Monroe, C.},
  journal = {Phys. Rev. X},
  volume = {5},
  issue = {2},
  pages = {021026},
  numpages = {9},
  year = {2015},
  month = {Jun},
  publisher = {American Physical Society},
  doi = {10.1103/PhysRevX.5.021026},
  url = {https://link.aps.org/doi/10.1103/PhysRevX.5.021026}
}

@article{NV_centers,
  title = {Clock transition by continuous dynamical decoupling of a three-level system},
  author = {Alexander Stark and Nati Aharon and Alexander Huck and Haitham A. R. El-Ella and Alex Retzker and 
Fedor Jelezko and Ulrik L. Andersen},
  journal = {Sci. Rep.},
  volume = {8},
  pages = {14807 },
  year = {2018}
}

@article{
    superconducting_artificial_atoms,
    author = {R. Bianchetti and S. Filipp and M. Baur and J. M. Fink and C. Lang and L. Steffen and M. Boissonneault and A. Blais and A. Wallraff},
    title = {Control and Tomography of a Three Level Superconducting Artificial Atom},
    journal = {Phys. Rev. Lett.},
    volume = {105},
    pages = {223601},
    year = {2010}
}

@article{qutip,
title = {\uppercase{Q}u\uppercase{T}i\uppercase{P} 2: A \uppercase{P}ython framework for the dynamics of open quantum systems},
journal = {Comput. Phys. Commun.},
volume = {184},
number = {4},
pages = {1234-1240},
year = {2013},
author = {J.R. Johansson and P.D. Nation and Franco Nori}
}

@Article{dirk2017,
author={Witthaut, Dirk
and Wimberger, Sandro
and Burioni, Raffaella
and Timme, Marc},
title={Classical synchronization indicates persistent entanglement in isolated quantum systems},
journal={Nat. Commun.},
year={2017},
month={Apr},
day={12},
volume={8},
number={1},
pages={14829},
issn={2041-1723},
doi={10.1038/ncomm}
}

@article{Zambrini2013,
  title = {Spontaneous synchronization and quantum correlation dynamics of open spin systems},
  author = {Giorgi, G. L. and Plastina, F. and Francica, G. and Zambrini, R.},
  journal = {Phys. Rev. A},
  volume = {88},
  issue = {4},
  pages = {042115},
  numpages = {9},
  year = {2013},
  month = {Oct},
  publisher = {American Physical Society},
  doi = {10.1103/PhysRevA.88.042115},
  url = {https://link.aps.org/doi/10.1103/PhysRevA.88.042115}
}

@Article{Zambrini_corr,
author={Manzano, Gonzalo
and Galve, Fernando
and Giorgi, Gian Luca
and Hern{\'a}ndez-Garc{\'i}a, Emilio
and Zambrini, Roberta},
title={Synchronization, quantum correlations and entanglement in oscillator networks},
journal={Sci. Rep.},
year={2013},
month={Mar},
day={14},
volume={3},
number={1},
pages={1439},
issn={2045-2322},
doi={10.1038/srep01439},
url={https://doi.org/10.1038/srep01439}
}

@article{Restrepo_dipoles,
doi = {10.1088/1367-2630/17/8/083063},
url = {https://doi.org/10.1088/1367-2630/17/8/083063},
year = {2015},
month = {sep},
publisher = {IOP Publishing},
volume = {17},
number = {8},
pages = {083063},
author = {Zhu, B and Schachenmayer, J and Xu, M and Herrera, F and Restrepo, J G and Holland, M J and Rey, A M},
title = {Synchronization of interacting quantum dipoles},
journal = {New J. Phys.}
}

@article{Bianconi_review,
doi = {10.1088/1751-8121/ad41a6},
url = {https://doi.org/10.1088/1751-8121/ad41a6},
year = {2024},
month = {may},
publisher = {IOP Publishing},
volume = {57},
number = {23},
pages = {233001},
author = {Nokkala, Johannes and Piilo, Jyrki and Bianconi, Ginestra},
title = {Complex quantum networks: a topical review},
journal = {J. Phys. A}
}

@article{csb_molnar,
year = {2020},
volume = {16},
pages = {351–356},
author = {Ferenc Molnar and Takashi Nishikawa and Adilson E. Motter},
title = {Network experiment demonstrates converse symmetry breaking},
journal = {Nat. Phys.}
}

@article{rajroy_motter,
  title = {Topological Control of Synchronization Patterns: Trading Symmetry for Stability},
  author = {Hart, Joseph D. and Zhang, Yuanzhao and Roy, Rajarshi and Motter, Adilson E.},
  journal = {Phys. Rev. Lett.},
  volume = {122},
  issue = {5},
  pages = {058301},
  numpages = {6},
  year = {2019},
  month = {Feb},
  publisher = {American Physical Society},
  doi = {10.1103/PhysRevLett.122.058301},
  url = {https://link.aps.org/doi/10.1103/PhysRevLett.122.058301}
}

@article{coherent_kiss_motter,
  title = {Coherent Dynamics Enhanced by Uncorrelated Noise},
  author = {Nicolaou, Zachary G. and Sebek, Michael and Kiss, Istv\'an Z. and Motter, Adilson E.},
  journal = {Phys. Rev. Lett.},
  volume = {125},
  issue = {9},
  pages = {094101},
  numpages = {6},
  year = {2020},
  month = {Aug},
  publisher = {American Physical Society},
  doi = {10.1103/PhysRevLett.125.094101},
  url = {https://link.aps.org/doi/10.1103/PhysRevLett.125.094101}
}

@article{enhance_sync_noise,
  title = {Enhancing Synchronization by Optimal Correlated Noise},
  author = {Martineau, Sherwood and Saffold, Tim and Chang, Timothy T. and Ronellenfitsch, Henrik},
  journal = {Phys. Rev. Lett.},
  volume = {128},
  issue = {9},
  pages = {098301},
  numpages = {7},
  year = {2022},
  month = {Mar},
  publisher = {American Physical Society},
  doi = {10.1103/PhysRevLett.128.098301},
  url = {https://link.aps.org/doi/10.1103/PhysRevLett.128.098301}
}

@article{csb_perego,
  title = {Synchronization and amplification enabled by diversity in nonlinear optical systems and the analogy with converse symmetry breaking for coupled oscillators},
  author = {Perego, Auro M.},
  journal = {Phys. Rev. A},
  volume = {106},
  issue = {3},
  pages = {L031505},
  numpages = {6},
  year = {2022},
  month = {Sep},
  publisher = {American Physical Society},
  doi = {10.1103/PhysRevA.106.L031505},
  url = {https://link.aps.org/doi/10.1103/PhysRevA.106.L031505}
}

@article{IBMQspin1,
  title={Quantum synchronization on the {IBM Q} system},
  author={Koppenh{\"o}fer, Martin and Bruder, Christoph and Roulet, Alexandre},
  journal={Phys. Rev. Res.},
  volume={2},
  number={2},
  pages={023026},
  year={2020},
  publisher={APS}
}

@misc{supp,
author = {},
title = {},
howpublished = {},
year = {},
note = 
{See Supplemental Material at [URL], which includes Refs.~\cite{atomic_coherent_states, glauber_spin_p, carmichael, Schwinger_rep, side_band_rev}, 
for details on the extremal eigenvalues of the spin-variance operator, the convergence timescale, the $P$-distribution and semiclassical model, pairwise synchronization measures, and the implementation of dissipation via sideband cooling.}
}

@article{atomic_coherent_states,
  title = {Atomic Coherent States in Quantum Optics},
  author = {Arecchi, F. T. and Courtens, Eric and Gilmore, Robert and Thomas, Harry},
  journal = {Phys. Rev. A},
  volume = {6},
  issue = {6},
  pages = {2211--2237},
  numpages = {0},
  year = {1972},
  month = {Dec},
  publisher = {American Physical Society},
  doi = {10.1103/PhysRevA.6.2211},
  url = {https://link.aps.org/doi/10.1103/PhysRevA.6.2211}
}

@article{glauber_spin_p,
  title={Superradiant pulses and directed angular momentum states},
  author={Glauber, Roy J and Haake, Fritz},
  journal={Phys. Rev. A},
  volume={13},
  number={1},
  pages={357},
  year={1976},
  publisher={APS}
}

@BOOK{carmichael,
   author = "H. J. Carmichael",
   title = "\uppercase{S}tatistical \uppercase{M}ethods in \uppercase{Q}uantum \uppercase{O}ptics 2: \uppercase{N}on-\uppercase{C}lassical \uppercase{F}ields",
   publisher = "Springer",
   address = "Berlin",
   year = "2008",
}

@article{Schwinger_rep,
year = {2011},
volume = {44},
pages = {065305},
author = {Ng, R and Sørensen, E S},
title = {Exact real-time dynamics of quantum spin systems using the positive-\uppercase{P} representation},
journal = {J. Phys. A: Math. Theor.},
}

@article{side_band_rev,
  title = {Quantum dynamics of single trapped ions},
  author = {Leibfried, D. and Blatt, R. and Monroe, C. and Wineland, D.},
  journal = {Rev. Mod. Phys.},
  volume = {75},
  pages = {281--324},
  numpages = {0},
  year = {2003},
}

@misc{github_spincsb,
    title = {},
    howpublished = {The results were obtained using the \uppercase{P}ython library \uppercase{Q}u\uppercase{T}i\uppercase{P} \cite{qutip}. Custom code is available at our \text{GitHub} repository: \url{https://github.com/dpbuser/spinCSBcodes.git/}}
}

@article{asymm_induced_qsync,
  title = {Quantum synchronization in a four-node optomechanical system via modulation asymmetry and nonlinear asymmetry},
  author = {Yang, W. D. and Sun, J. T. and Liu, H. D.},
  journal = {Phys. Rev. A},
  volume = {113},
  issue = {2},
  pages = {023503},
  numpages = {10},
  year = {2026},
  month = {Feb},
  publisher = {American Physical Society}
}

@article{dahmen2026how,
title = {How heterogeneity shapes dynamics and computation in the brain},
journal = {Neuron},
volume = {114},
number = {5},
pages = {804-819},
year = {2026},
issn = {0896-6273},
author = {David Dahmen and Axel Hutt and Giacomo Indiveri and Ann Kennedy and Jeremie Lefebvre and Luca Mazzucato and Adilson E. Motter and Rishikesh Narayanan and Melika Payvand and Henrike Planert and Richard Gast}
}

@article{sagues2007spatiotemporal,
title={Spatiotemporal order out of noise},
author={Sagu{\'e}s, Francesc and Sancho, Jos{\'e} M and Garc{\'\i}a-Ojalvo, Jordi},
journal={Rev. Mod. Phys.},
volume={79},
number={3},
pages={829--882},
year={2007},
publisher={APS}
}

@article{nair2021using,
  title = {Using Disorder to Overcome Disorder: A Mechanism for Frequency and Phase Synchronization of Diode Laser Arrays},
  author = {Nair, N. and Hu, K. and Berrill, M. and Wiesenfeld, K. and Braiman, Y.},
  journal = {Phys. Rev. Lett.},
  volume = {127},
  issue = {17},
  pages = {173901},
  numpages = {7},
  year = {2021},
  publisher = {American Physical Society}
}

@article{braiman1995taming,
  title = {Taming spatiotemporal chaos with disorder},
  author = {Y. Braiman and John F. Lindner and William L. Ditto },
  journal = {Nature},
  volume = {378},
  pages = {465–467},
  year = {1995}
}

@article{brandt2006synchronization,
  title = {Synchronization from Disordered Driving Forces in Arrays of Coupled Oscillators},
  author = {Brandt, Sebastian F. and Dellen, Babette K. and Wessel, Ralf},
  journal = {Phys. Rev. Lett.},
  volume = {96},
  issue = {3},
  pages = {034104},
  numpages = {4},
  year = {2006},
  publisher = {American Physical Society}
}

@article{tessone2006diversity,
  title = {Diversity-Induced Resonance},
  author = {Tessone, Claudio J. and Mirasso, Claudio R. and Toral, Ra\'ul and Gunton, James D.},
  journal = {Phys. Rev. Lett.},
  volume = {97},
  issue = {19},
  pages = {194101},
  numpages = {4},
  year = {2006},
  publisher = {American Physical Society}
}

@article{montaseri2016diversity,
  title = {Diversity of coupled oscillators can enhance their synchronization},
  author = {Montaseri, Ghazal and Meyer-Hermann, Michael},
  journal = {Phys. Rev. E},
  volume = {94},
  issue = {4},
  pages = {042213},
  numpages = {10},
  year = {2016},
  publisher = {American Physical Society}
}

@article{punetha2022how,
  title = {How clock heterogeneity affects synchronization and can enhance stability},
  author = {Punetha, Nirmal and Wetzel, Lucas},
  journal = {Phys. Rev. E},
  volume = {106},
  issue = {5},
  pages = {054216},
  numpages = {14},
  year = {2022},
  publisher = {American Physical Society}
}

@article{maltsev2022functional,
title={Functional heterogeneity of cell populations increases robustness of pacemaker function in a numerical model of the sinoatrial node tissue},
author={Maltsev, Alexander V and Stern, Michael D and Lakatta, Edward G and Maltsev, Victor A},
journal={Front. Physiol.},
volume={13},
pages={845634},
year={2022},
publisher={Frontiers Media SA}
}

@article{zhang2022isochronous,
  title={Isochronous synchronization induced by topological heterogeneity in semiconductor laser networks},
  author={Zhang, Liyue and Pan, Wei and Yan, Lianshan and Luo, Bin and Zou, Xihua and Li, Songsui},
  journal={Opt. Laser Technol.},
  volume={153},
  pages={108243},
  year={2022},
  publisher={Elsevier}
}

@article{arthur2025optimal,
  title = {Optimal flock formation induced by agent heterogeneity},
  author = {Arthur N. Montanari and Ana Elisa D. Barioni and Chao Duan and Adilson E. Motter},
  journal = {Nat. Commun.},
  volume = {16},
  issue = {1},
  pages = {9626},
  year = {2025}
}

@article{yang2022emergent,
  title={Emergent microrobotic oscillators via asymmetry-induced order},
  author={Yang, Jing Fan and Berrueta, Thomas A and Brooks, Ada M and Liu, Albert Tianxiang and Zhang, Ge and Gonzalez-Medrano, David and Yang, Sungyun and Koman, Volodymyr B and Chvykov, Pavel and LeMar, Lexy N and others},
  journal={Nat. Commun.},
  volume={13},
  number={1},
  pages={5734},
  year={2022},
  publisher={Nature Publishing Group UK London}
}

@article{gaussian_tomography,
  title = {Efficient Method for Computing the Maximum-Likelihood Quantum State from Measurements with Additive \uppercase{G}aussian Noise},
  author = {Smolin, John A. and Gambetta, Jay M. and Smith, Graeme},
  journal = {Phys. Rev. Lett.},
  volume = {108},
  pages = {070502},
  numpages = {4},
  year = {2012},
}

%============================================================
%=============== Supplemental Material ======================
%============================================================
\let\addcontentsline\oldaddcontentsline
\onecolumngrid
\newpage

\renewcommand{\thepage}{\arabic{page}}
\setcounter{page}{1}

\renewcommand{\theequation}{S\arabic{equation}}

\renewcommand{\thesection}{S\arabic{section}}

\renewcommand{\thefigure}{S\arabic{figure}}
\setcounter{figure}{0}

{\centering
{\Large\sc{Supplemental Material}}\\
\vspace{1mm}
for \\
\vspace{1.2mm}
{\Large Quantum manifestations of converse symmetry breaking}\\
\vspace{1mm}
Daniel P. Buser, Biswabibek Bandyopadhyay, and Adilson E. Motter\\
}

\bigskip

\startcontents[sm]

%\tableofcontents
\titlecontents{section}
  [1.5em]
  {\vspace{10pt}}
  {\contentsmargin{0em}\thecontentslabel\hspace{1em}}  
  {\hspace{0em}}
  {\normalfont\hfill\thecontentspage}

\printcontents[sm]{}{1}

\bigskip

\newcounter{lasteq}
%\newpage

\section{Extremal eigenvalues of the spin variance operator $R$}
\label{sec:R_min_max}
\setcounter{equation}{0}

The spin variance operator is defined as 
\begin{equation}
    R=\frac{1}{N}\sum_{n=1}^N ||\vec{S}_n-\overline{\vec{S}}||^2,
    \label{eq:R_measure}
\end{equation}
where $\overline{\vec{S}}=\frac{1}{N}\sum_{n=1}^N\vec{S}_n=\frac{\vec{S}_{\rm tot}}{N}$. Using $||\vec{A}||^2=\vec{A}\cdot\vec{A}$, we find a new form of $R$ as follows:
\begin{align}
    \nonumber
    R&=\frac{1}{N}\sum_{n=1}^N(\vec{S}_n-\overline{\vec{S}})\cdot(\vec{S}_n-\overline{\vec{S}}) \\ 
    \nonumber
    &=\frac{1}{N}\sum_{n=1}^N(\vec{S}_n^2-2\vec{S}_n\cdot\overline{\vec{S}}+\overline{\vec{S}}^2) \\
    \nonumber
    &=\frac{1}{N}\Big[\sum_{n=1}^N\vec{S}_n^2-2\sum_{n=1}^N \vec{S}_n\cdot\overline{\vec{S}} + N\overline{\vec{S}}^2\Big] \\
    \nonumber
    &=\frac{1}{N}\Big[\sum_{n=1}^N S(S+1)\mathds{1}-2\sum_{n=1}^N\vec{S}_n\cdot\frac{1}{N}\sum_{m=1}^N\vec{S}_m+\frac{N}{N^2}(\sum_{n=1}^N\vec{S}_n)^2 \Big] \\
    \nonumber
    &=\frac{1}{N}\Big[N S(S+1)\mathds{1}-\frac{2}{N}\sum_{n,m}\vec{S}_n\cdot \vec{S}_m+\frac{1}{N}\vec{S}_{\rm tot}^2 \Big] \\
    \nonumber
    &=\frac{1}{N}\Big[N S(S+1)\mathds{1}-\frac{2}{N}(\sum_{n=1}^N\vec{S}_n)^2+\frac{1}{N}\vec{S}_{\rm tot}^2 \Big] \\
    \nonumber
    &=\frac{1}{N}\Big[N S(S+1)\mathds{1}-\frac{2}{N}\vec{S}_{\rm tot}^2+\frac{1}{N}\vec{S}_{\rm tot}^2 \Big] \\
    \nonumber
    &=S(S+1)\mathds{1}-\frac{1}{N^2}\vec{S}_{\rm tot}^2,
\end{align}
where each particle has fixed spin $S$ and $\mathds{1}$ is the identity operator on the full $N$-spin Hilbert space. As $[R,\vec{S}_{\rm tot}^2]=0$, we can choose a standard angular momentum eigenbasis for the total spin of the composite spin system ($\ket{S_{\rm tot}, M}$) that simultaneously diagonalizes both operators. Here, $S_{\rm tot}$ is the total spin quantum number and $M$ is the total magnetization with $\vec{S}_{\rm tot}^2\ket{S_{\rm tot},M}=S_{\rm tot}(S_{\rm tot}+1)\ket{S_{\rm tot},M}$ and $\vec{S}_{\rm tot}^z\ket{S_{\rm tot},M}=M\ket{S_{\rm tot},M}$. Operating $R$ on $\ket{S_{\rm tot},M}$, we obtain
\begin{align}
    \nonumber
    R\ket{S_{\rm tot},M}&=\Big[S(S+1)\mathds{1}-\frac{1}{N^2}\vec{S}_{\rm tot}^2 \Big]\ket{S_{\rm tot},M} \\
    &=\Big[S(S+1)-\frac{S_{\rm tot}(S_{\rm tot}+1)}{N^2} \Big]\ket{S_{\rm tot},M},
\end{align}
where $\lambda(S_{\rm tot})=S(S+1)-\frac{S_{\rm tot}(S_{\rm tot}+1)}{N^2}$ is the eigenvalue of $R$ associated with the eigenstate $\ket{S_{\rm tot},M}$. Since $\lambda(S_{\rm tot})$ decreases monotonically with $S_{\rm tot}$, we can write $\lambda(S_{\rm tot}^{\rm max})=\lambda_{\rm min}$ and $\lambda(S_{\rm tot}^{\rm min})=\lambda_{\rm max}$. For a composite system of $N$ spins of magnitude $S$, the total angular momentum $S_{\rm tot}$ satisfies
\begin{align}
    S_{\rm tot}^{\max} &= NS,\\
S_{\rm tot}^{\min} &=
\begin{cases}
0, & NS \in \mathbb{N},\\
\tfrac{1}{2}, & NS \in \mathbb{N}+\tfrac{1}{2},
\end{cases}
\end{align}
where $\mathbb{N}$ is the set of natural numbers. Finally, we obtain
\begin{align}
    \lambda_{\rm min} &= S(S+1)-\frac{NS(NS+1)}{N^2} 
    =\frac{S(N-1)}{N},\\
\lambda_{\rm max} &=
\begin{cases}
S(S+1), & NS \in \mathbb{N},\\
S(S+1)-\frac{3}{4N^2}, & NS \in \mathbb{N}+\tfrac{1}{2}.
\end{cases}
\label{eq:min_max}
\end{align}

The minimum is achieved when all spins are aligned in the same direction, such as when all $N$ spins are in the maximal eigenstate of $S^z$. This maximum is achieved, for example, in the case of an antisymmetric superposition of three spin-1 particles in coherent states $2\pi/3$ apart in any plane, including the center of the $(\theta, \phi)$ sphere.
\setcounter{lasteq}{\value{equation}}

\section{Convergence timescale} \label{sec:timescale}
\setcounter{equation}{\value{lasteq}}

\begin{figure}
     \centering
     \includegraphics[scale=0.5]{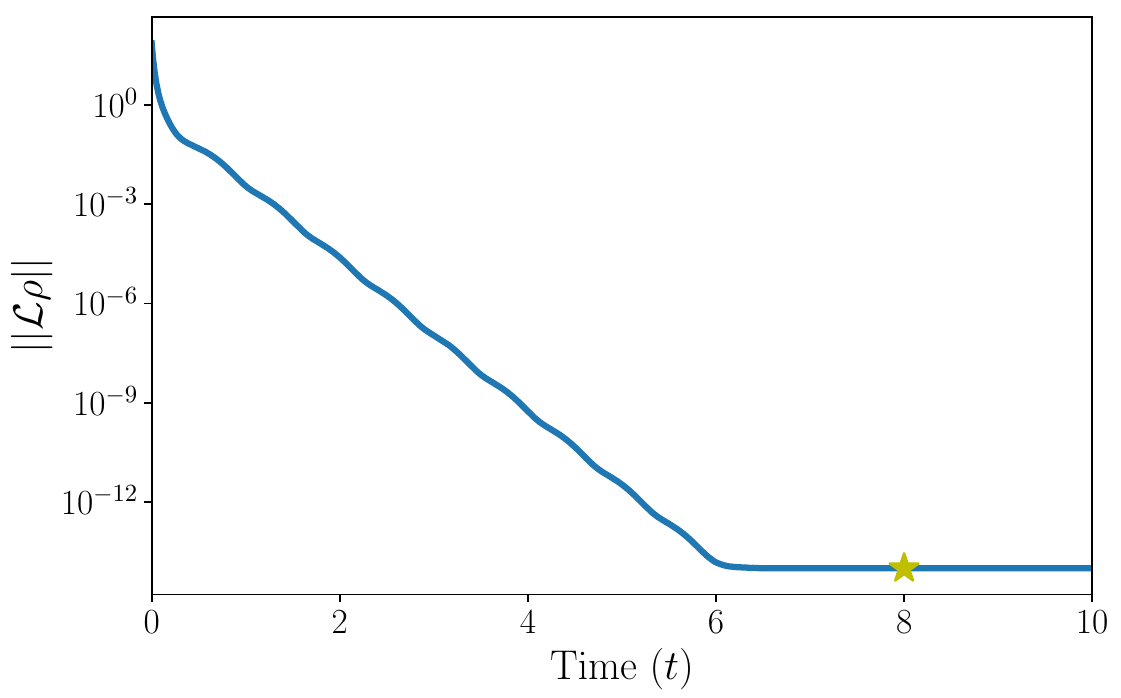}
     \caption{Approach to equilibrium, quantified by the Frobenius norm of $\mathcal{L}\rho(t)$ as a function of time. The simulation is performed for the three-spin system with $S=8$ shown in Fig.~\ref{fig:RMS}, starting from the initial state  $|S,S \rangle \langle S, S|^{\otimes 3}$. The function $||\mathcal{L}\rho(t)||$ follows an exponential decay before saturating at $\approx 10^{-13}$, which is effectively zero within numerical precision.
    The yellow star indicates the cutoff time used in the simulations.}
     \label{fig:convergence}
 \end{figure}

All simulations of the quantum model were performed with the initial condition  $\rho = |S,S\rangle \langle S, S|^{\otimes 3}$. The convergence of this state to the steady state of Eq.~\eqref{eq:quantum_model} is shown in Fig.~\ref{fig:convergence}. The plotted quantity is the Frobenius norm of $\mathcal{L}\rho$, the Liouvillian acting on the state $\rho$ as
defined by Eq.~\eqref{eq:quantum_model} 
in the form $\dot{\rho}=\mathcal{L}\rho$. As $\rho$ evolves, it approaches the steady state $\rho_{\rm ss}$ for which $\mathcal{L}\rho$ is zero. The yellow star indicates the cutoff time used in our simulations for the exact quantum model, corresponding to approximately $8$ units of time, by which the system has reached its steady state within numerical precision. The quantum model plots of Fig.~\ref{fig:RMS} were obtained using this cutoff. The semiclassical plots of Fig.~\ref{fig:RMS} were averaged over 780 units after a transient time of 20 units.
\setcounter{lasteq}{\value{equation}}

\section{$P$-distribution and the semiclassical model}
\label{sec:p_distribution}
\setcounter{equation}{\value{lasteq}}

We now derive the Langevin equations for a semiclassical description of the spin vdP oscillator system, which also offers an insightful alternative route to obtaining the classical equations. 
This is done through the intermediate use of the $P$-representation, which is obtained by expanding the density matrix $\rho$ in terms of spin coherent states
\begin{equation}\label{eq:p_rep}
\rho = \int_{\partial B^3} P(\theta, \phi) |\theta, \phi \rangle \langle \theta, \phi | d\Omega, \;\;\;\;\;\;\;\;\;\;\;\; |\theta, \phi \rangle = \cos\left(\frac{\theta}{2}\right)^{2S}e^{\tan\left(\frac{\theta}{2}\right)e^{{\rm i}\,\phi}S_-}| S, S \rangle,
\end{equation}
where $\partial B^3$ denotes the boundary of the three-dimensional unit ball. A more complete description of spin coherent states and their uses may be found in~\cite{atomic_coherent_states, glauber_spin_p}.

The master equation in Eq.~\eqref{eq:quantum_model} can be converted into a partial differential equation using the following single-spin relations, which employ the use of the variable $u = \tan(\theta/2) \exp({\rm i}\,\phi)$ with no subscript or overhead arrow:
\begin{subequations}\label{eq:conversion_rules}
    \begin{equation}
        S^z |u\rangle\langle u| = \left( S\frac{1 - u^* u}{1 + u^*u} - u\frac{\partial}{\partial u}\right) |u\rangle\langle u|,
    \end{equation}
    \begin{equation}
        S^- |u\rangle\langle u| = \left( 2Su^* \left(1+u^* u\right)^{-1} + \frac{\partial}{\partial u}\right) |u\rangle\langle u|,
    \end{equation}
    \begin{equation}
        S^+ |u\rangle\langle u| = \left( 2Su \left(1 + u^*u\right)^{-1} - u^2 \frac{\partial}{\partial u} \right) |u\rangle\langle u|,
    \end{equation}
\end{subequations}
Before proceeding, we first introduce the following notation for the multi-spin coherent state: 
    \begin{equation}
    |\vec{u}\rangle\langle\vec{u}| = \bigotimes_{n=1}^N |u_n\rangle\langle u_n|.
    \end{equation}
The terms of Eq.~\eqref{eq:quantum_model} can now be converted into their corresponding $P$-representation as follows:
\begin{subequations}
    \begin{equation}
        \big[S^z_n, |\vec{u}\rangle\langle\vec{u}|\big] = \left( -u_n\frac{\partial}{\partial u_n} - \text{c.c.} \right)|\vec{u}\rangle\langle\vec{u}|,
    \end{equation}
    \begin{equation}
        \mathcal{D}[S_n^-]|\vec{u}\rangle\langle\vec{u}| = \left( (S+1) u_n\frac{\partial}{\partial u_n} +\frac{1}{2}u_n^2\frac{\partial^2}{\partial u_n^2} + \frac{1}{2}\frac{\partial^2}{\partial u_n^*\partial u_n} + \text{c.c.} \right)|\vec{u}\rangle\langle\vec{u}|,
    \end{equation}
    \begin{gather}
        \mathcal{D}[S^{+^2}_n]|\vec{u}\rangle\langle\vec{u}| = \Biggl(\frac{2S|u_n|^2u_n\left( (1-2S)|u_n|^2 + 2S(1-2S) \right)}{(1 + |u_n|^2)^2} \frac{\partial}{\partial u_n}\notag \\ + \frac{Su_n^2\left( (2S-1)|u_n|^4 + 4(2S - 1)|u_n|^2 - (2S-1) \right)}{(1 + |u_n|^2)^2} \frac{\partial^2}{\partial u_n^2} + 2|u_n|^6\frac{\left( 4S(S-1) - 2S|u_n|^2 (1 + |u|^2)^2\right)}{\left( |u_n|^2 + 1\right)^2}\frac{\partial^2}{\partial u_n^*\partial u_n}\notag \\  + \frac{u_n^3\left( 4S - (2S+1)(1+|u_n|^2) \right)}{1 + |u_n|^2}\frac{\partial^3}{\partial u_n^3}\notag + \frac{2|u_n|^6u_n(|u_n|^2 -2S + 1)}{|u_n|^2 + 1}\frac{\partial^3}{\partial u_n^*\partial u_n^2} \\ - \frac{1}{2}u_n^2\frac{\partial^4}{\partial u_n^4} + \frac{1}{2}|u_n|^8\frac{\partial^4}{\partial {u_n^{*}}^2\partial u_n^2} + \text{c.c.} \Biggl)|\vec{u}\rangle\langle\vec{u}|,
    \end{gather}
    \begin{align}
        \sum_{n,m=1}^N C^{h+}_{nm}\mathcal{D}[S_n^-, S_m^+]|\vec{u}\rangle\langle\vec{u}| =& \Biggl(\sum_{n,m=1}^N \frac{S(C_{nm}^{h+}u_m + C_{nm}^{h+^*}u_n^2u_m^*)}{1 + u_m^*u_m}\frac{\partial}{\partial u_n}\notag \\+& \frac{1}{2}C_{nm}^{h+^*}u_n^2\frac{\partial^2}{\partial u_n\partial u_m} + \frac{1}{2}C_{nm}^{h+}\frac{\partial^2}{\partial u_n\partial u_m^*} + \text{c.c.}\Biggl)|\vec{u}\rangle\langle\vec{u}|,
    \end{align}
    \begin{align}
        \sum_{n,m=1}^N -C^{h-^*}_{nm}\mathcal{D}[S_n^+, S_m^-]|\vec{u}\rangle\langle\vec{u}| =& \Biggl(\sum_{n,m=1}^N \frac{S(C_{nm}^{h-}u_m + C_{nm}^{h-^*}u_n^2u_m^*)}{1 + u_m^*u_m}\frac{\partial}{\partial u_n}\notag \\-& \frac{1}{2}C_{nm}^{h-^*}u_n^2\frac{\partial^2}{\partial u_n\partial u_m} - \frac{1}{2}C_{nm}^{h-^*}u_n^2u_m^{*^2}\frac{\partial^2}{\partial u_n\partial u_m^*} + \text{c.c.}\Biggl)|\vec{u}\rangle\langle\vec{u}|,
    \end{align}
    \begin{equation}
        \frac{1}{2}\sum_{n,m=1}^N C_{nm}^a [S_n^-S_m^+, |\vec{u}\rangle\langle\vec{u}|] = \left(\sum_{n,m=1}^N \frac{S(C_{nm}^{a}u_m + C_{nm}^{a^*}u_n^2u_m^*)}{1 + u_m^*u_m} \frac{\partial}{\partial u_n} + \frac{1}{2}C_{nm}^{a^*} u_n^2 \frac{\partial^2}{\partial u_n \partial u_m} + \text{c.c.}\right)|\vec{u}\rangle\langle\vec{u}|.
    \end{equation}
\end{subequations}
After substituting Eq.~\eqref{eq:p_rep} into Eq.~\eqref{eq:quantum_model}, one can replace each term in the master equation by its corresponding differential operator given above. Upon completing this and integrating by parts, one obtains
\begin{gather}
    \frac{\partial P}{\partial t} = \Biggl[ \sum_{n=1}^N -\frac{\partial}{\partial u_n}\Biggl( {\rm i}\,\omega_nu_n + \left( \gamma_\text{g}\frac{(S+1)}{S} + \frac{1}{S}\sum_{m=1}^N C_{mm}^{h+} \right)u_n - \gamma_\text{d}\frac{2S-1}{S^2}u_n|u_n|^2\frac{|u_n|^2 + 2S}{(1 + |u_n|^2)^2} + \sum_{m=1}^N \frac{C_{nm}u_m + C_{nm}^*u_n^2u_m^*}{1 + |u_m|^2} \Biggl)\notag \\ + \frac{1}{2}\sum_{n,m=1}^N \frac{\partial^2}{\partial u_n\partial u_m}\frac{1}{S}\left( \left( \delta_{nm}\left(\gamma_\text{g} + \gamma_\text{d} \frac{2S-1}{S}\frac{|u_n|^4 + 4|u_n|^2 - 1}{(1 + |u_n|^2)^2}\right) + (C_{nm}^{h+^*} - C_{nm}^{h-^*} + C_{nm}^{a^*}) \right)u_n^2 \right)\notag \\ + \frac{1}{2}\sum_{n,m=1}^N \frac{\partial^2}{\partial u_n\partial u_m^*}\frac{1}{S}\left( \delta_{nm}\left(\gamma_\text{g} + \gamma_\text{d}\frac{|u_n|^6}{S^2} \left( \frac{4S(S-1) - 2S|u_n|^2}{(1+|u_n|^2)^2} + 1 \right) \right) + C_{nm}^{h+^*} - C_{nm}^{h-^*}u_n^2u_m^{*^2} \right)\notag \\ + \sum_{n=1}^N -\frac{\partial^3}{\partial u_n^3} \left( \gamma_\text{d}\frac{u_n^3}{S^3}\left( \frac{4S}{1 + |u_n|^2} - 2S - 1 \right) \right) - \sum_{n=1}^N \frac{\partial^3}{\partial u_n^*\partial u_n^2}\left( \frac{\gamma_\text{d}}{S^3}|u_n|^6u_n\frac{1-2S+|u_n|^2}{1 + |u_n|^2} \right)\notag \\ - \sum_{n=1}^N \frac{\partial^4}{\partial u_n^4}\left( \frac{\gamma_\text{d}}{4S^3}u_n^2 \right) + \sum_{n=1}^N \frac{\partial^4}{\partial {u_n^{*}}^2\partial u_n^2}\left( \frac{\gamma_\text{d}}{4S^3}|u_n|^8 \right) + {\rm c.c.} \Biggl] P(\vec{u}, \vec{u}^*).
    \label{eq:P_equation}
\end{gather}
This equation is not yet in Fokker-Planck form as it contains derivatives of order higher than two. After dropping all higher-order terms, a valid Fokker-Planck equation will be obtained. One peculiarity remains, however, in that the diffusion matrix obtained from the second-order terms is generally not positive semidefinite. If one wishes to obtain valid Langevin equations, there are two approaches to remedy this situation. The first is to simply drop problematic terms from the diffusion matrix to make it positive semidefinite. The second is to re-derive the equation in the positive $P$-representation. This involves decomposing the density matrix in terms of non-diagonal projectors $|u\rangle\langle\beta|/\langle\beta|u\rangle$ rather than the diagonal projectors in Eq.~\eqref{eq:p_rep}. The added degrees of freedom in this decomposition will guarantee a positive semidefinite diffusion that can be modeled in a Langevin equation. Although it will not be investigated further here, what one will find is that the phase space equation in the positive $P$-representation is identical to Eq.~\eqref{eq:P_equation} with $u_n^*$ replaced with the complex variable $\beta_n$. 
Further details can be found in Ref.~\cite{carmichael}.

To make the diffusion matrix corresponding to Eq.~\eqref{eq:P_equation} positive semidefinite, all off-diagonal terms contained in $\partial_{u_n}\partial_{u_m}$ and $\partial_{u_n}\partial_{u_m^*}$ must be dropped. After this, one may obtain the equivalent Langevin equations:
\begin{gather}
    \dot{u}_n = \Biggl[{\rm i}\,\omega_nu_n + \left( \gamma_\text{g} \left(\frac{S+1}{S}\right) + \frac{1}{S}\sum_{m=1}^NC_{mm}^{h+} \right)u_n - \gamma_\text{d} \frac{2S-1}{S^2}\left|u_n\right|^2u_n \frac{|u_n|^2 + 2S}{(1 + |u_n|^2)^2} + \sum_{m=1}^N \frac{C_{nm}u_m + C_{nm}^* u_n^2 u_m^*}{1 + |u_m|^2}\Biggl]\notag \\
    + \left[ \frac{1}{2S} \gamma_\text{g} + \frac{1}{2} \gamma_\text{d} \frac{|u_n|^6}{S^2} \left(\frac{4S(S-1) - 2S|u_n|^2}{(1+|u_n|^2)^2} + \frac{1}{S}\right)\right]^{\frac{1}{2}}\left(\xi(t)_n^R + {\rm i}\,\xi(t)_n^I\right),
    \label{eq:langevin_full}
\end{gather}
which corresponds to Eq.~\eqref{eq:P_equation} when the diagonal terms $\partial_{u_n}\partial_{u_n}$ are dropped in addition to those already neglected above. Upon taking the limit of Eq.~\eqref{eq:langevin_full} as $S\to\infty$, the noise terms vanish, and the following classical equation is obtained:
\begin{gather}
    \dot{u}_n = {\rm i}\,\omega_nu_n + \gamma_\text{g} u_n - 4\gamma_\text{d} \frac{\left|u_n\right|^2u_n}{(1 + |u_n|^2)^2} + \sum_{m=1}^N \frac{C_{nm}u_m + C_{nm}^* u_n^2 u_m^*}{1 + |u_m|^2}.
\end{gather}
The relations $\theta_n = 2\arctan\left(|u_n|\right)$ and $\phi_n = \arg(u_n)$, together with the chain rule equations $\dot{\theta}_n = \dot{u}_n(\partial_{u_n}\theta_n) + \text{c.c.}$ and $\dot{\phi}_n = \dot{u}_n(\partial_{u_n}\phi_n) + \text{c.c.}$,
yield the time dependence of the angular coordinates. 
Substituting $u_n=\tan(\theta_n/2)e^{{\rm i}\,\phi_n}$
then recovers the classical Eq.~\eqref{eq:classical_model_coupled}.

\setcounter{lasteq}{\value{equation}}

\section{Pairwise synchronization measures}
\label{sec:pairwise_r}
\setcounter{equation}{\value{lasteq}}

Consider the spin variance measure $R$ as defined in Eq.~\eqref{eq:R_measure} in the case where the sum runs over two out of the three spins. The three possible pairwise choices for $R$ are then
\begin{subequations}
    \begin{equation}
        R_{12} = S(S+1) - \frac{1}{4}(\vec{S}_1 + \vec{S}_2)^2 = S(S+1) - \frac{1}{4} (2S(S+1) + 2\vec{S}_1\cdot\vec{S}_2),
    \end{equation}
    \vspace{-4mm}
    \begin{equation}
        R_{13} = S(S+1) - \frac{1}{4}(\vec{S}_1 + \vec{S}_3)^2 = S(S+1) - \frac{1}{4} (2S(S+1) + 2\vec{S}_1\cdot\vec{S}_3),
    \end{equation}
    \vspace{-4mm}
    \begin{equation}
        R_{23} = S(S+1) - \frac{1}{4}(\vec{S_2} + \vec{S_3})^2 = S(S+1) - \frac{1}{4} (2S(S+1) + 2\vec{S}_2\cdot\vec{S}_3).
    \end{equation}
\end{subequations}
Summing the three measures together and multiplying by $4/9$ then yields:
\begin{align}
    \frac{4}{9}(R_{12} + R_{13} + R_{23}) &= \frac{4}{9}\left( \frac{3}{2}S(S+1) - \frac{1}{4}(2\vec{S}_1 \cdot \vec{S}_2 + 2\vec{S}_1\cdot\vec{S}_3 + 2\vec{S}_2\cdot\vec{S}_3) \right)=\notag \\ &=\frac{2}{3}S(S+1) - \frac{1}{9}(2\vec{S}_1 \cdot \vec{S}_2 + 2\vec{S}_1\cdot\vec{S}_3 + 2\vec{S}_2\cdot\vec{S}_3)= S(S+1) - \frac{1}{9}(\vec{S}_1 + \vec{S}_2 + \vec{S}_3)^2 = R,
\end{align}
which is the spin variance operator for all three spins. For integer $S$, as considered throughout, the sum of the  $r_{ij}$ measures for $(i,j) = (1,2)$, $(1,3)$, and $(2,3)$ is
\begin{equation}
    \sum_{(i,j)}r_{ij} = \sum_{(i,j)}\frac{\langle R_{ij}\rangle - \lambda_{(i,j) \text{min}}}{\lambda_{(i,j) \text{max}} - \lambda_{(i,j) \text{min}}} = \frac{\frac{9}{4}\langle R\rangle-\frac{3S}{2}}{S(S+1) - \frac{S}{2}} = \frac{\frac{9}{4}\langle R\rangle - \frac{9}{4}(\frac{2S}{3})}{S(S+1) - \frac{3}{4}(\frac{2S}{3})},
\end{equation}
where $\lambda_{(i,j)\text{min}}$ and $\lambda_{(i,j)\text{max}}$ are the pairwise counterparts to $\lambda_{\text{min}}$ and $\lambda_{\text{max}}$, respectively. Multiplying the above expression by the scaling factor
\begin{equation}
    q = \frac{4\big[S(S+1)-\frac{3}{4}(\frac{2S}{3})\big]}{9\big[S(S+1)-(\frac{2S}{3})\big]}
    \label{eq:scaling_factor}
\end{equation}
finally gives
\begin{equation}
    \frac{4\big(S(S+1)-\frac{3}{4}(\frac{2S}{3})\big)}{9\big(S(S+1)-(\frac{2S}{3})\big)}\sum_{(i,j)}r_{ij} =\frac{\langle R\rangle - (\frac{2S}{3})}{S(S+1) - (\frac{2S}{3})} = \frac{\langle R\rangle - \lambda_{\text{min}}}{\lambda_{\text{max}} - \lambda_{\text{min}}} = r.
\end{equation}
For the case of $S=6$, $q$ evaluates to 26/57 and can be applied to the pixelwise sum of the three plots in Fig.~\ref{fig:RMS_vs_phase}(b) to obtain Fig.~\ref{fig:RMS_vs_phase}(d).

\begin{figure*}
    \includegraphics[scale=0.28]{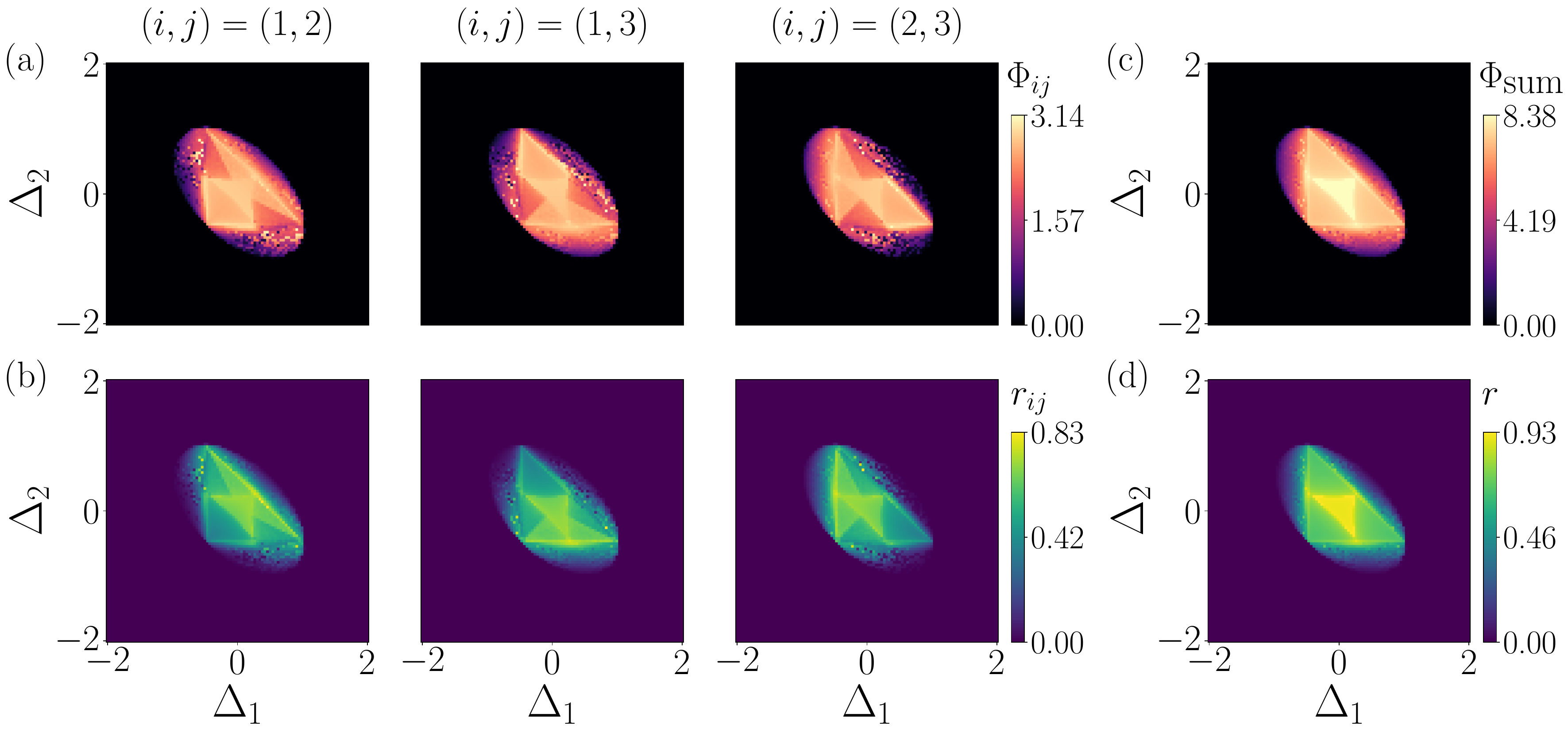}
\caption{Pairwise dynamics for the classical counterpart  ($S\to\infty$) of the 3-spin system in Fig.~\ref{fig:RMS_vs_phase}.
Plots in panel (a) are the effective phase-locking angles $\Phi_{ij}$ as defined in Eq.~\eqref{eq:rel_phase}. Panel (c) is the pixelwise sum of the three plots in (a). Panel (d) can be calculated by using Eq.~\eqref{eq:rms} or, equivalently, by a rescaled pixelwise sum of the three plots in panel (b). The regularization is set to $\epsilon=0.1$, and the other parameters are the same as in
Fig.~\ref{fig:RMS_vs_phase}.}
\label{fig:classical_pairwise}
\end{figure*}

In the classical limit, the values of MTLE can be compared to the $r$-parameter and phase-locking angle, which can be derived from the following equations:
\begin{equation}
    \label{eq:r-classical}
    \Tilde{r} = \left(\frac{N-1}{N}\right)-\frac{2}{N^2}\sum_{n=1}^N \sum_{m>n}^N \frac{2(u_n u_m^* + u_n^* u_m) + (1 - |u_n|^2)(1 - |u_m|^2)}{(1 + |u_n|^2)(1 + |u_m|^2)},
\end{equation}
\begin{equation}
    \label{eq:phi_classical}
    \Tilde{\Phi}_{ij} = \arg(u_i) - \arg(u_j).
\end{equation}
Equation~\eqref{eq:r-classical} is derived by substituting the expectation value $\langle \vec{u} | R | \vec{u} \rangle$ into Eq.~\eqref{eq:rms}, where $|\vec{u}\rangle = \bigotimes_{n=1}^N|u_n\rangle$, $|u_n\rangle = e^{-{\rm i}\,\phi_nS^z_n}e^{-{\rm i}\,\theta_nS^y_n}|S, S\rangle$, and $u_n=\tan(\theta_n/2)e^{{\rm i}\,\phi_n}$, and then taking the classical limit of $r$ as $S\to \infty$. Similarly, Eq.~\eqref{eq:phi_classical} is derived as the classical limit of Eq.~\eqref{eq:rel_phase} in which the phase distribution $S_{ij}(\phi)$ becomes sharply peaked about the classical phase difference. 
We now use Eqs.~\eqref{eq:r-classical} and \eqref{eq:phi_classical} to calculate steady-state values of $r$ and $\Phi_{ij}$ for the classical counterparts of Figs.~\ref{fig:RMS} and \ref{fig:RMS_vs_phase}, in which $\langle R\rangle$ was calculated against the steady-state solution of
Eq.~\eqref{eq:master_equation}.
To obtain the steady-state values $r$ and $\Phi_{ij}$ in the classical limit, both $\Tilde{r}$ and $\Tilde{\Phi}_{ij}$ must be time averaged according to the dynamics of Eq.~\eqref{subeq:langevin_eq} in the $S\to\infty$ limit for a sufficiently long time. The classical plots of Fig.~\ref{fig:classical_pairwise} were calculated according to this time averaging procedure
with a small value of the regularization parameter, $\epsilon=0.1$.
Figure~\ref{fig:classical_pairwise}(d)
was calculated by applying a pixelwise sum of the plots in Fig.~\ref{fig:classical_pairwise}(b)
rescaled by a factor of $q = 4/9$, as given by Eq.~\eqref{eq:scaling_factor} in the limit $S\to \infty$.
The region of instability shown in the center of these plots (where the measures are nonzero) agrees with the region of instability shown in the MTLE plot of Fig.~\ref{fig:mtle} for $A=0$.

\setcounter{lasteq}{\value{equation}}
\section{Implementation of dissipation via sideband cooling}
\label{sec:experiment}
\setcounter{equation}{\value{lasteq}}

The main challenge in experimentally implementing the system described by Eq.~\eqref{eq:master_equation} lies in the 
complex nature of the dissipators and coupling terms. 
Rather than realizing the model directly using coupled spins, one can map this system onto the vibrational modes of trapped ions via the Schwinger representation of spin operators \cite{Schwinger_rep}:
\begin{equation}
    S^+ \to ba^\dag,\;\;\;\;\;\;\;\;\; S^- \to b^\dag a,\;\;\;\;\;\;\;\;\; S^z \to \frac{1}{2}(a^\dag a - b^\dag b),
\end{equation}
where $[a,a]=[b,b]=[a,b]=[a,b^\dag]=0$ and $[a,a^\dag]=[b,b^\dag]=1$.
It can be shown that all commutators are preserved by this mapping. The spin states map to
\begin{equation}
    |S,m\rangle \to |S+m\rangle\otimes |S-m\rangle,
\end{equation}
where $a^\dag a |S+m,S-m\rangle = (S+m)|S+m,S-m\rangle$ and $b^\dag b|m,S-m\rangle = (S-m)|S+m,S-m\rangle$.
Transforming the master equation in Eq.~\eqref{eq:master_equation} according to this prescription yields:
\begin{align}
    \nonumber
    \dot{\rho} &= \sum_{n=1}^N -\frac{1}{2}{\rm i}\,\omega[a_n^\dag a_n - b_n^\dag b_n, \rho] + \frac{\gamma_g}{S}\mathcal{D}[b_n^\dag a_n]\rho + \frac{\gamma_d}{2S^3}\mathcal{D}[b_n^2 (a_n^\dag)^2]\rho \\&-\frac{{\rm i}\,}{2S} \lambda_n^a \left[\left(\sum_{l=1}^N u^a_{ln} b_l^\dag a_l\right)\left(\sum_{l=1}^N u^a_{ln} b_l^\dag a_l\right)^\dag, \rho\right] +\frac{\lambda_n^{h+}}{S}\mathcal{D}\left[\sum_{l=1}^N u^h_{ln} b_l^\dag a_l\right]\rho - \frac{\lambda_n^{h-}}{S}\mathcal{D}\left[\sum_{l=1}^N (u^h_{ln})^*b_l a_l^\dag\right]\rho, 
    \label{eq:schwinger_rep}
\end{align}
where $\lambda^{h+}_n$ are the positive eigenvalues of $C^{h+}$, $\lambda^{h-}_n$ are the negative eigenvalues of $C^{h-}$, and $\lambda^a_n$ are the eigenvalues of ${\rm i}\,C^a$. Furthermore, $u^h_{ln}$ and $u^a_{ln}$ are the elements of the diagonalizing matrix for $C^h$ and ${\rm i}\,C^a$, respectively.
All Hamiltonian terms appearing in the Schwinger-transformed model given by Eq.~\eqref{eq:schwinger_rep}, as well as all additional Hamiltonians required to simulate the dissipative terms in this equation, are assumed to be experimentally implementable. 
The rest of this section will focus on how to implement the dissipators, which consist of a short-time application of the sideband-cooling unitary operation followed by a measurement on the auxiliary spin degree of freedom. 
The Hamiltonians to be directly implemented couple an auxiliary spin degree of freedom to an external field. Under the rotating-wave approximation, these Hamiltonians effectively become the sideband Hamiltonians.

Denoting $\vec{\sigma}=(\sigma^x, \sigma^y, \sigma^z)$ as the vector of Pauli matrices, the Hamiltonian for sideband cooling of a single mode is
\begin{align}
    H &= -\frac{g}{2} \vec{\sigma}\cdot \vec{E}\notag + H_0 \\ &=-\frac{{\rm i}\,g E_0}{2} \cos(kz-\omega_L t)(\sigma^- - \sigma^+) + H_0,
\end{align}
where $\sigma^\pm=\frac{1}{2}(\sigma^x\pm {\rm i}\,\sigma^y)$, $g$ is a coupling constant, $E_0$ is the electric field amplitude, $k$ is the wave number of the field, and $\omega_L$ is its angular frequency. It is also assumed that the electric field's polarization is in the $y$ direction. Additionally, the non-interacting Hamiltonian is
\begin{equation}
    H_{0} = \omega a^\dag a + \nu |\mathfrak{e}\rangle\langle \mathfrak{e} |,
    \label{eq:int_ham}
\end{equation}
where $\omega$ is the oscillator frequency and $\nu$ is the energy of the spin excited state $|\mathfrak{e}\rangle$ relative to the ground state $|\mathfrak{g}\rangle$. It can be shown that, in the interaction picture under the rotating-wave approximation, this reduces to the effective sideband Hamiltonian
\begin{equation}
    H_{\rm sb} = \frac{\Omega}{2}(O\sigma^+ + O^\dag \sigma^-), \label{eq:sb_ham}
\end{equation}
where $\Omega=-(1/2)gE_0 z_0 k$ with $z_0$ being the zero-point fluctuations of the vibrational mode $a$. We first restrict the operators $O$ to $a,a^2,\dots$ in the case of red sideband driving and $a^\dag,(a^\dag)^2,\dots$ in the case of blue sideband driving \cite{qsync_expt1, side_band_rev}. If this Hamiltonian is applied for a small time $\delta t$, the following is obtained:
\begin{align}
    \rho(t+\delta t) &= \text{Tr}_{\rm spin}\big(e^{-{\rm i}\, H_{\rm sb}\delta t} \rho \otimes |\mathfrak{g}\rangle\langle \mathfrak{g} | e^{{\rm i}\, H_{\rm sb}\delta t}\big)\notag \\&=\rho(t) + \Gamma \delta t O\rho(t)O^\dag - \frac{\Gamma}{2}\delta t\big\{ O^\dag O, \rho(t) \big\},
    \label{eq:sb_diss}
\end{align}
which is the dissipator in Lindblad form, where $\Gamma = (\Omega/2)^2 \delta t$ \cite{qsync_expt1}. Additionally, the final line of Eq.~\eqref{eq:sb_diss} is the result of a measurement on the auxiliary spin followed by a reset to the ground state $|\mathfrak{g}\rangle\langle \mathfrak{g} |$.
In practice, it is the stochastic Sch\"odinger equation that is being implemented rather than the Lindblad master equation.
In a single time step $\delta t$, the transition probability to the state $O|\Psi\rangle/\sqrt{\langle O^\dag O\rangle}$ is $\Gamma \delta t \langle O^\dag O\rangle$. The Lindblad form is only recovered as an average of the density matrices produced by many iterations of the stochastic Schr\"odinger equation.

Next, we set $O=b^\dag a$ to match the dissipator corresponding to linear gain in Eq.~\eqref{eq:schwinger_rep}. This can be done by applying two sideband drives to the modes $a$ and $b$, with the respective quantities and operators indexed by $1$ and $2$. 
The double sideband Hamiltonian can then be expressed by introducing the drive strengths $\Omega_1$ and $\Omega_2$, which induce two different transitions $(|\mathfrak{g}_1\rangle \leftrightarrow |\mathfrak{e}\rangle$ and $|\mathfrak{g}_2 \rangle \leftrightarrow |\mathfrak{e}\rangle)$, as
\begin{equation}
    H_{\rm dsb} = \frac{\Omega_1}{2}\big(O_1|\mathfrak{e}\rangle\langle \mathfrak{g}_1 | + O_1^\dag |\mathfrak{g}_1\rangle\langle \mathfrak{e} |\big) + \frac{\Omega_2}{2}\big(O_2 |\mathfrak{e}\rangle\langle \mathfrak{g}_2 | + O_2^\dag |\mathfrak{g}_2 \rangle\langle \mathfrak{e} |\big) + H_{\rm shift},
    \label{eq:dsb_ham}
\end{equation}
where $|\mathfrak{g}_1\rangle$ and $|\mathfrak{g}_2 \rangle$ are two ground states whose energy spacing is negligible compared to their spacing $\nu$ with the excited state $|\mathfrak{e}\rangle$. Since Eq.~\eqref{eq:dsb_ham} is the sum of two sideband Hamiltonians, the operators $O_1$ and $O_2$ are constrained to $O_1=a, a^\dag, a^2, (a^\dag)^2 ,...)$ and $O_2=b, b^\dag, b^2, (b^\dag)^2 ,...)$ for modes $a$ and $b$ such that $[a,a^\dag]=[b,b^\dag]=1$ and $[a,b]=0$. We now restore the non-interaction term of the spin contribution from Eq.~\eqref{eq:int_ham}, which gives each term a time-dependent prefactor:
\begin{equation}
    H_{\rm dsb}(t) = \nu |\mathfrak{e}\rangle\langle \mathfrak{e} | + \frac{\Omega_1}{2}\big(O_1 e^{-{\rm i}\,\nu t} |\mathfrak{e}\rangle\langle \mathfrak{g}_1 | + O_1^\dag e^{{\rm i}\,\nu t}|\mathfrak{g}_1\rangle\langle \mathfrak{e} |\big) + \frac{\Omega_2}{2}\big(O_2 e^{-{\rm i}\,\nu t} |\mathfrak{e}\rangle\langle \mathfrak{g}_2 | + O_2^\dag e^{{\rm i}\,\nu t} |\mathfrak{g}_2 \rangle\langle \mathfrak{e} |\big) + H_{\rm shift}.
\end{equation}
The Hamiltonian $H_{\rm shift}$ is yet to be determined and will be used to cancel undesirable terms. 

One can write the wavefunction as
\begin{equation}
    |\Psi\rangle = \sum_{\alpha,\beta=0}^\infty c_{\mathfrak{e},\alpha, \beta}|\mathfrak{e},\alpha,\beta\rangle + c_{\mathfrak{g}_1,\alpha,\beta}|\mathfrak{g}_1,\alpha,\beta\rangle + c_{\mathfrak{g}_2,\alpha,\beta}|\mathfrak{g}_2,\alpha,\beta \rangle,
    \label{eq:wavefunction}
\end{equation}
where $|\alpha\rangle$ and $|\beta\rangle$ are Fock states such that $a^\dag a|\alpha, \beta\rangle=\alpha|\alpha, \beta\rangle$ and $b^\dag b|\alpha, \beta\rangle=\beta|\alpha, \beta\rangle$. The dummy operators $O_1$ and $O_2$ act only on the two Fock subspaces such that $O_1$ acts on the first and $O_2$ on the second ($O_1|\alpha,\beta\rangle = O_1|\alpha\rangle \otimes |\beta\rangle$ and $O_2|\alpha,\beta\rangle = |\alpha\rangle \otimes O_2|\beta\rangle$). To obtain an effective Hamiltonian that will implement dissipation of the form $b^\dag a$, one must make the adiabatic approximation by eliminating the excited state coefficient $c_{\mathfrak{e},n,m}$ for all $n,m$ from the time evolution of $c_{\mathfrak{g}_1,n,m}$ and $c_{\mathfrak{g}_2,n,m}$. The coefficients $c_{\mathfrak{e},n,m}$ evolve in time as
\begin{equation}
    {\rm i}\,\dot{c}_{\mathfrak{e},n,m} = \nu c_{\mathfrak{e},n,m} + \frac{\Omega_1}{2} e^{-{\rm i}\,\nu t}\Big(\sum_\alpha c_{\mathfrak{g}_1,\alpha, m}\langle n|O_1|\alpha \rangle\Big)+\frac{\Omega_2}{2}e^{-{\rm i}\,\nu t}\Big(\sum_\beta c_{\mathfrak{g}_2,n,\beta}\langle m|O_2|\beta\rangle\Big) + \langle e,n,m| H_\text{shift}|\Psi\rangle.
\end{equation}
For sufficiently large $\nu$, we can use the approximation $\dot{c}_{\mathfrak{e},n,m}=0$. Additionally, we impose the constraint that $H_\text{shift}|e,n,m\rangle = 0$ for all $n,m$. Solving for $c_{\mathfrak{e},n,m}$ gives
\begin{equation}
    c_{\mathfrak{e},n,m} = -\frac{1}{\nu}\Big[ \frac{\Omega_1}{2} e^{-{\rm i}\,\nu t}\Big(\sum_\alpha c_{\mathfrak{g}_1,\alpha, m}\langle n|O_1|\alpha \rangle\Big)+\frac{\Omega_2}{2}e^{-{\rm i}\,\nu t}\Big(\sum_\beta c_{\mathfrak{g}_2,n,\beta}\langle m|O_2|\beta\rangle\Big)\Big].
\end{equation}
Now, by substituting the expression for $c_{\mathfrak{e}, n, m}$ above and solving for $\dot{c}_{\mathfrak{g}_1,n,m}$ and $\dot{c}_{\mathfrak{g}_2,n,m}$, we obtain
\begin{align}
    {\rm i}\,\dot{c}_{\mathfrak{g}_1,n,m} &= \frac{\Omega_1}{2} e^{{\rm i}\,\nu t} \Big(\sum_\alpha c_{\mathfrak{e},\alpha,m} \langle n|O_1^\dag |\alpha\rangle\Big) + \langle \mathfrak{g}_1,n,m|H_{\text{shift}}|\Psi\rangle \\\notag 
    &= -\frac{1}{\nu}\Big[ \frac{\Omega_1\Omega_2}{4}\Big(\sum_{\alpha,\beta} c_{\mathfrak{g}_2,\alpha,\beta}\langle n| O_1^\dag |\alpha\rangle\langle m|O_2|\beta\rangle\Big) + \frac{\Omega_1^2}{4}\Big(\sum_{\alpha} c_{\mathfrak{g}_1, \alpha,m}\langle n|O_1^\dag O_1|\alpha\rangle\Big) \Big] + \langle \mathfrak{g}_1,n,m|H_{\text{shift}}|\Psi\rangle, \notag\\
    {\rm i}\,\dot{c}_{\mathfrak{g}_2,n,m} &= \frac{\Omega_2}{2}e^{{\rm i}\,\nu t}\Big(\sum_{\beta}c_{\mathfrak{e},n,\beta}\langle m|O_2^\dag|\beta\rangle\Big) + \langle \mathfrak{g}_2,n,m|H_{\text{shift}}|\Psi\rangle \notag \\ &= -\frac{1}{\nu}\Big[\frac{\Omega_1\Omega_2}{4}\Big(\sum_{\alpha,\beta}c_{\mathfrak{g}_1,\alpha ,\beta}\langle n| O_1|\alpha\rangle\langle m|O^\dag_2|\beta\rangle\Big)+\frac{\Omega_2^2}{4}\Big(\sum_\beta c_{\mathfrak{g}_2,n,\beta}\langle m|O_2^\dag O_2|\beta\rangle\Big)\Big]+ \langle \mathfrak{g}_2,n,m|H_{\text{shift}}|\Psi\rangle.
\end{align}
This leads to the  effective Hamiltonian
\begin{equation}
    H_{\rm eff} = -\frac{\Omega_1\Omega_2}{4\nu} \big(O_1 O_2^\dag |\mathfrak{g}_2 \rangle\langle \mathfrak{g}_1 | + O_1^\dag O_2 |\mathfrak{g}_1\rangle\langle \mathfrak{g}_2 |\big) - \frac{\Omega_1^2}{4\nu}O_1^\dag O_1 |\mathfrak{g}_1\rangle\langle \mathfrak{g}_1 | - \frac{\Omega_2^2}{4\nu}O_2^\dag O_2 |\mathfrak{g}_2 \rangle\langle \mathfrak{g}_2 | + H_{\rm shift},
\end{equation}
which can be verified by confirming ${\rm i}\,c_{\mathfrak{g}_1,n,m}=\langle \mathfrak{g}_1,n,m|H_\text{eff}|\Psi\rangle$ and ${\rm i}\,c_{\mathfrak{g}_2,n,m}=\langle \mathfrak{g}_2,n,m|H_\text{eff}|\Psi\rangle$. Choosing $H_{\rm shift}$ to cancel the final two terms then gives
\begin{equation}
    H_{\rm eff} = -\frac{\Omega_1\Omega_2}{4\nu} \big(O_1 O_2^\dag |\mathfrak{g}_2 \rangle\langle \mathfrak{g}_1 | + O_1^\dag O_2 |\mathfrak{g}_1\rangle\langle \mathfrak{g}_2 |\big)
    \label{eq:dsb_final}.
\end{equation} Next, set $O$ in Eq.~\eqref{eq:sb_ham} to $O_1 O_2^\dag$. For $O_1 = a$ and $O_2 = b$, this gives us a dissipator from Eq.~\eqref{eq:sb_diss} whose argument is $b^\dag a$ as desired, which is  the linear gain from Eq.~\eqref{eq:schwinger_rep}. Similarly, the nonlinear damping may be derived by setting $O_1=(a^\dag)^2$ and $O_2=(b^\dag)^2$.

It is straightforward to extend the result of Eq.~\eqref{eq:dsb_final} to the case of more than two modes in the wavefunction, which is necessary for implementing the dissipators in Eq.~\eqref{eq:schwinger_rep}. Specifically, this means the wavefunction in Eq.~\eqref{eq:wavefunction} now takes the following form:
\begin{equation}
    |\Psi\rangle = \sum_{\vec{\alpha},\vec{\beta}=0}^\infty c_{\mathfrak{e},\vec{\alpha}, \vec{\beta}}|\mathfrak{e},\vec{\alpha},\vec{\beta}\rangle + c_{\mathfrak{g}_1,\vec{\alpha},\vec{\beta}}|\mathfrak{g}_1,\vec{\alpha},\vec{\beta}\rangle + c_{\mathfrak{g}_2,\vec{\alpha},\vec{\beta}}|\mathfrak{g}_2,\vec{\alpha},\vec{\beta} \rangle,
\end{equation}
where $\vec{\alpha}=(\alpha_1,...,\alpha_N)$, $a_l^\dag a_l|\mathfrak{s},\vec{\alpha},\vec{\beta}\rangle=\alpha_l|\mathfrak{s},\vec{\alpha},\vec{\beta}\rangle$ for $\mathfrak{s}=\mathfrak{e}$, $\mathfrak{g}_1$,  $\mathfrak{g_2}$, and likewise for the $\vec{\beta}$ modes. Starting with the Hamiltonian from Eq.~\eqref{eq:dsb_ham} where now $O_1=a_l, a_l^\dag, a_l^2, (a_l^\dag)^2 ,...$ and $O_2=b_l, b_l^\dag, b_l^2, (b_l^\dag)^2 ,...$ for $l=1,...,N$, one will then arrive at a multimode effective Hamiltonian that has the same form as the Hamiltonian in Eq.~\eqref{eq:dsb_final}.

Finally, one must deal with the coupling terms within the dissipators of Eq.~\eqref{eq:schwinger_rep}. This requires variable coupling constants $g_n$, variable electric field strengths $E_{ln}$, and variable phase shifts $\phi_{ln}$. The coupling factor $g$ now has a subscript $n$ in order to implement the dissipator prefactors $\lambda_n^{h\pm}$ in Eq.~\eqref{eq:schwinger_rep}. Likewise the strengths $E_{ln}$ and phases $e^{{\rm i}\,\phi_{ln}}$ are used to implement the diagonalizing matrix elements $u_{ln}$. Consider the Hamiltonian
\begin{align}
    H_{ln} &= -\frac{g_n}{2} \vec{E}_{ln} \cdot \vec{\sigma} + H_0\notag \\
    &= -\frac{g_n}{2} E_{ln}\cos(kz_l-\omega_L t + \phi_{ln})\hat{y} \cdot \vec{\sigma} +  H_0,
    \label{eq:Eln}
\end{align}
where once again we have $E_{ln}$ polarized in the $y$-direction. Taking $\sigma^y$ to equal ${\rm i}\,(|\mathfrak{g}\rangle\langle \mathfrak{e} | - |\mathfrak{e}\rangle\langle \mathfrak{g} |)$ and substituting this into Eq.~\eqref{eq:Eln} gives
\begin{equation}
    H_{ln}=-g_n \frac{{\rm i}\, E_{ln}}{4} \big(e^{{\rm i}\,(kz_l+\phi_{ln})} e^{-{\rm i}\,\omega_L t} + e^{-{\rm i}\,(k z_l +\phi_{ln})}e^{{\rm i}\,\omega_L t}\big)(|\mathfrak{g}\rangle\langle \mathfrak{e} | - |\mathfrak{e}\rangle\langle \mathfrak{g} |) + H_0.
\end{equation}
Moving into the interaction picture with the non-interacting Hamiltonian $H_{0} = \omega a_l^\dag a_l + \nu |\mathfrak{e}\rangle\langle \mathfrak{e} |$, one gets
\begin{align}
\label{eq:int_pic_H}
    H_{ln}(t) &= - g_n \frac{{\rm i}\,E_{ln}}{4} \big(e^{{\rm i}\,(kz_l(t)+\phi_{ln})} e^{-{\rm i}\,\omega_L t} + e^{-{\rm i}\,(k z_l(t) +\phi_{ln})}e^{{\rm i}\,\omega_L t}\big)\big(e^{-{\rm i}\,\nu t}|\mathfrak{g}\rangle\langle \mathfrak{e} | - e^{{\rm i}\,\nu t}|\mathfrak{e}\rangle\langle \mathfrak{g} |\big)\notag \\ &\approx g_n\frac{E_{ln}}{4}\big({\rm i}\,e^{{\rm i}\,(kz_l(t)+\phi_{ln})}e^{{\rm i}\,\delta t}|\mathfrak{e}\rangle\langle \mathfrak{g} | + {\rm h.c.}\big),
\end{align}
where the $\delta = \nu-\omega_L$ is a detuning parameter. Here, we have dropped the fast rotating terms to obtain the approximate form in 
Eq.~\eqref{eq:int_pic_H}. Now, expanding the exponentials $e^{{\rm i}\,kz_l(t)}$ to first order, we obtain
\begin{align}
    H_{ln}(t) &\approx g_n \frac{E_{ln}}{4}\left[{\rm i}\,\big(1 + {\rm i}\,k z_0(a_le^{-{\rm i}\,\omega t}+a_l^\dag e^{{\rm i}\,\omega t})\big)e^{{\rm i}\,\delta t}e^{{\rm i}\,\phi_{ln}}|\mathfrak{e}\rangle\langle \mathfrak{g} | + {\rm h.c.}\right]\notag \\
    &\approx - g_n\frac{E_{ln}}{4}(\eta a_l e^{{\rm i}\,\phi_{ln}}|\mathfrak{e}\rangle\langle \mathfrak{g} | + {\rm h.c.}),
\end{align}
where we have used $z_l(t)=z_0(a_le^{-{\rm i}\,\omega t}+a_l^\dag e^{{\rm i}\,\omega t})$, and the Lamb-Dick parameter is defined as $\eta = z_0 k$. The second approximation in this equation follows from setting $\delta=\omega$ and neglecting the fast rotating terms.
Finally, defining $\sqrt{u_{ln}} = E_{ln}e^{{\rm i}\,\phi_{ln}}/4$, one obtains
\begin{equation}
    H_{ln} = -g_n\eta \sqrt{u_{ln}} a_l |\mathfrak{e}\rangle\langle \mathfrak{g} | + {\rm h.c.}
\end{equation}

We can add the second sideband drive for the mode $b_l$ as follows:
\begin{equation}
    H_{ln}^{\rm dsb} = -g_n\eta(\sqrt{u_{ln}}a_l |\mathfrak{e}\rangle\langle \mathfrak{g}_1 | - \sqrt{u_{nl}^*}b_l |\mathfrak{e}\rangle\langle \mathfrak{g}_2 | + {\rm h.c.}) + H_{\rm shift},
    \label{eq:coupling_sb}
\end{equation}
where once again we choose $H_{\rm shift}$ to cancel undesirable terms.
From this form one has that $O_1 = \sqrt{u_{ln}}a_l$ and $O_2 = \sqrt{u_{ln}^*}b_l$, and $\Omega_1 = -\Omega_2 = -2g_n\eta $. According to the forms of Eqs.~\eqref{eq:dsb_ham} and \eqref{eq:dsb_final}, this leads to an effective Hamiltonian of the form
\begin{equation}
    H^{\rm eff}_{ln} = \frac{g_n^2\eta^2}{\nu}(u_{ln}a_l b_l^\dag |\mathfrak{g}_2\rangle\langle \mathfrak{g}_1 | + {\rm h.c.}).
\end{equation}
Next, if these Hamiltonians over various $l$ are applied in rapid succession such that $\delta t$ is sufficiently small, the following approximations hold:
\begin{align}
    \exp(-{\rm i}\,H^{\rm eff}_{1n}\delta t)...\exp(-{\rm i}\,H^{\rm eff}_{Nn}\delta t)&\approx(1-{\rm i}\,H^{\rm eff}_{1n}\delta t)...(1-{\rm i}\,H^{\rm eff}_{Nn}\delta t)\notag\\\nonumber
    &\approx\Big(1-{\rm i}\,\sum_{l=1}^N H^{\rm eff}_{ln}\delta t\Big)\\ \nonumber
    &\approx \exp\Big(-{\rm i}\,\sum_{l=1}^N H^{\rm eff}_{ln}\delta t\Big) \\
    &= \exp(-{\rm i}\,H^{\rm sum}_n \delta t).
\end{align}
Thus, the Hamiltonian $H^{\rm sum}_n$ takes the form
\begin{equation}
  H^{\rm sum}_n = \frac{g_n^2\eta^2}{\nu}\Big(\sum_{l=1}^N u_{ln}a_l b_l^\dag \sigma^+ + {\rm h.c.}\Big). 
\end{equation}
Substituting $(g_n^2 \eta^2/\nu)$ for $\Omega/2$ and $\sum_l u_{ln}a_l b_l^\dag $ for $O$ in Eq.~\eqref{eq:sb_ham} leads to the following time evolution for sufficiently small $\delta t$:
\begin{equation}
    \rho(t+\delta t) = \rho(t) + \frac{\lambda_n}{S} \mathcal{D}\Big[\sum_l u_{ln}a_l b_l^\dag\Big]\rho(t) \delta t,
\end{equation}
where $\lambda_n/S = (g_n^4\eta^4/\nu^2)\delta t$. At this point, one would then measure the auxiliary levels to advance the stochastic Schr\"odinger equation by $\delta t$. 

Repeating this whole process over all values of $n$, we obtain
\begin{equation}
    \rho(t+\delta t) = \rho(t) + \sum_{n=1}^N\frac{\lambda_n}{S} \mathcal{D}\Big[\sum_{l=1}^N u_{ln}a_l b_l^\dag\Big]\rho(t)\delta t,
\end{equation}
which corresponds to the first sum of coupling dissipators in the index $n$ with strengths $\lambda^{h+}_n$ in Eq.~\eqref{eq:schwinger_rep} by identifying $\lambda_n$ with $\lambda_n^{h+}$. To obtain the second set of coupling dissipators, which have strengths $\lambda^{h-}_n$, one simply modifies Eq.~\eqref{eq:coupling_sb} by taking $O_1 = \eta_n\sqrt{u_{ln}^*}a_l^\dag$ and $O_2 = \eta_n\sqrt{u_{ln}}b_l^\dag$. Repeating the process again, but taking instead $-\lambda_n^{h-}/S = (g_n^4\eta^4/\nu^2)\delta t$, one obtains
\begin{equation}
    \rho(t+\delta t) = \rho(t) - \sum_{n=1}^N\frac{\lambda_n^{h-}}{S}\mathcal{D}\Big[\sum_{l=1}^N u^*_{ln}a_l^\dag b_l\Big]\rho(t)\delta t,
\end{equation}
which corresponds to the second sum of coupling dissipators in Eq.~\eqref{eq:schwinger_rep}.

\end{document}